\documentclass[12pt]{article}
\usepackage{amsmath}
\usepackage{graphicx}
\usepackage{enumerate}
\usepackage{natbib}
\usepackage{url} 
\usepackage[hypertexnames=false]{hyperref}
\usepackage{tabularx, booktabs}
\usepackage{lscape}

\usepackage{subfig}

\newcommand{\blind}{1}

\usepackage{amsthm, amssymb, mathrsfs, mathtools, bm, bbm, dutchcal, dsfont}
\usepackage{setspace}
\usepackage{comment}
\usepackage{xcolor}

\usepackage{float}
\usepackage{fancyhdr}
\usepackage{arydshln}
\usepackage{placeins}
\usepackage{enumitem, tablefootnote}
\usepackage{lineno}

\newcommand{\kv}{k}
\newcommand{\kvp}{k^{\prime}}

\newcommand{\thb}{\bm{\theta}}

\newcommand{\mub}{\bm{\mu}}
\newcommand{\Sigb}{\bm{\Sigma}}
\newcommand{\thbs}{\bm{\theta^\ast}}
\newcommand{\thbt}{\bm{\tilde{\theta}}}
\newcommand{\mubs}{\bm{\mu^\ast}}
\newcommand{\Sigbs}{\bm{\Sigma^\ast}}
\newcommand{\xb}{\bm{x}}
\newcommand{\eb}{\bm{e}}
\newcommand{\yb}{\bm{y}}

\newcommand{\Qb}{\bm{Q}}
\newcommand{\Rb}{\bm{R}}
\newcommand{\Rbt}{\bm{\tilde{R}}}
\newcommand{\Sb}{\bm{S}}
\newcommand{\Sbt}{\bm{\tilde{S}}}
\newcommand{\Db}{\bm{D}}
\newcommand{\Pb}{\bm{P}}
\newcommand{\Vb}{\bm{V}}
\newcommand{\Vbar}{\bar{V}}

\newcommand{\Zb}{\bm{Z}}
\newcommand{\Zhv}{\bm{\hat{Z}}_{v}}
\newcommand{\Zhe}{\bm{\hat{Z}}_{e}}
\newcommand{\Zbv}{\bm{Z}_{v}}

\newcommand{\ii}{^{-i}}
\newcommand{\gn}{G^{(N)}}
\newcommand{\gt}{\widetilde{G_{\mbox{\tiny VZ}}}} 
\newcommand{\gtn}[1]{\widetilde{G_{\mbox{\tiny VZ}}^{(#1)}}}
\newcommand{\gtnn}{\gtn{N}}

\newcommand{\tildegn}{\widetilde{\gn}}
\newcommand{\gnt}{\tildegn}
\newcommand{\ginf}{G^{(\infty)}}

\newcommand{\Norm}{\text{N}}
\newcommand{\NIW}{\text{NIW}}

\newcommand{\Dir}{\text{Dir}}
\newcommand{\GEM}{\text{GEM}}
\newcommand{\DM}{\text{DM}}

\newcommand{\Ind}{\mathbbm{1}}
\renewcommand{\Re}{\mathbbm{R}}
\newcommand{\N}{\mathbbm{N}}
\newcommand{\iidsim}{\overset{\text{iid}}{\sim}}
\newcommand{\indsim}{\overset{\text{ind}}{\sim}}
\newcommand{\EPPF}{\mathrm{EPPF}}
\newcommand{\fEPPF}{\mathrm{fEPPF}}

\newcommand{\Zv}{\bm{Z_{v}}}
\newcommand{\Ze}{\bm{Z_{e}}}
\newcommand{\Tr}{E_{N}}

\newcommand{\Mep}{M_{e}^{+}}

\usepackage[normalem]{ulem}

\newtheorem{Thm}{\underline{\bf Theorem}}

\newtheorem*{Proof*}{Proof}

\newtheorem{Prop}{\underline{\bf Proposition}}

\newtheorem{Cor}{\underline{\bf Corollary}}
\newtheorem{Exa}{Example}

\begin{document}

\def\spacingset#1{\renewcommand{\baselinestretch}%
{#1}\small\normalsize}\spacingset{1}


\if1\blind
{
\begin{center}
	{\LARGE{\bf \mbox{Graph-Aligned Random Partition Model} (GARP)
	}}
\end{center}
\vskip 2mm
\begin{center}
	\small
	Giovanni Rebaudo$^{a}$ (giovanni.rebaudo@unito.it) \\
	Peter M\"uller$^{b}$ (pmueller@math.utexas.edu) \\
	\vskip 3mm
     $^{a}$
     University of Torino, IT
  	\vskip 4pt 
	$^{b}$
 University of Texas at Austin, USA\\ 
\end{center}
} \fi

\if0\blind
{
  \bigskip
  \bigskip
  \bigskip
  \begin{center}
    {\LARGE\bf 	Graph-Aligned Random Partition Model (GARP)}
\end{center}
  \medskip
} \fi

\bigskip
\begin{abstract}
Bayesian nonparametric mixtures and random partition models are powerful tools for probabilistic clustering.
However, standard independent mixture models can be restrictive in some applications such as inference on cell lineage due to the biological relations of the clusters.
The increasing availability of large genomic data requires new statistical tools to perform model-based clustering and infer the relationship between homogeneous subgroups of units.
Motivated by single-cell RNA data we develop a novel dependent mixture model to jointly perform cluster analysis and align the clusters on a graph.
Our flexible graph-aligned random partition model (GARP) exploits Gibbs-type priors as building blocks, allowing us to derive analytical results for the probability mass function (pmf) on the graph-aligned random partition.
We derive a generalization of the Chinese restaurant process from the pmf and a related efficient and neat MCMC algorithm to implement Bayesian inference.
We illustrate posterior inference under the GARP using single-cell RNA-seq data from mice stem cells.
We further investigate the performance of the model in recovering the underlying clustering structure as well as the underlying graph by means of simulation studies. 
\end{abstract}

\noindent%
Dependent Mixture Model, 
Exchangeability, 
Single-Cell RNA

\spacingset{1.2}
\section{Introduction} \label{sec: intro}
We introduce a graph-aligned random partition model with one set of clusters being identified as vertices of a graph and other clusters being interpreted as edges between those. 
The model construction is motivated by the increasing availability of genomic data that requires new statistical tools to perform inference and uncertainty quantification on homogeneous subgroups of units (e.g., single-cells) and hypothesized relationships between the subgroups (e.g., transitions between the subgroups).
In the present article, we deal with single-cell RNA sequencing experiments (scRNA-seq) that provide an unprecedented opportunity to study cellular heterogeneity and the evolution of complex tissues. 
The interest is to identify the main homogeneous cell subpopulations (i.e., clusters) in terms of gene expressions and jointly infer transitions of cells between these.

Dirichlet process (DP) mixtures \citep{lo1984class} are well-established Bayesian nonparametric (BNP) models to infer homogeneous subgroups of observations via probabilistic clustering.
However, the law of the random partition induced by the DP, related to the so-called Chinese restaurant process (CRP), is controlled by a single parameter.  
This leaves DP mixture models too restrictive for many applications and several alternative models were introduced in the literature to allow more flexible clustering.
This includes the symmetric finite Dirichlet prior \citep{green2001modelling}, the Pitman-Yor process (PYP) \citep{pitman1997two}, the normalized inverse Gaussian (NIG) \citep{lijoi2005hierarchical}, the normalized generalized gamma process (NGGP) \citep{lijoi2007controlling}, mixture of finite mixtures (MFM) \citep{nobile1994bayesian, richardson1997bayesian, nobile2007bayesian, miller2018mixture} and
the mixture of DP (MDP) models \citep{antoniak1974mixtures}. 
All these belong to the wider family of Gibbs-type priors \citep{gnedin2006exchangeable} that can be seen as a natural, flexible generalization of the DP \citep{deblasi2015gibbs}. 

However, Gibbs-type processes entail independent cluster-specific parameters not allowing us to infer the relationship between clusters as needed in our motivating example. 
Recently, repulsive priors that allow for dependent cluster-specific parameters were successfully introduced to favor more parsimonious and well-separated clusters \citep{petralia2012repulsive, xu2016bayesian, beraha2022mcmc}.
Repulsive mixtures introduce (negative) dependence between cluster-specific values to better separate clusters. 
However, these models still stop short of inferring a biological relationship between the clusters, such as aligning the clusters on a graph, as desired in our framework.

\begin{figure}[ht]
    \centering
    \includegraphics[width=0.5\linewidth]{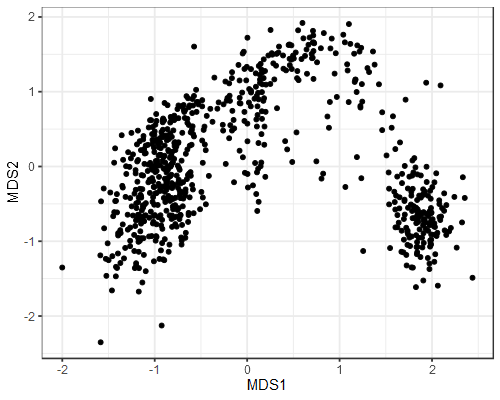}
    \caption{Two-dimensional representation of genetic expressions of the RNA mice single-cells data.
    \label{fig: mice data}}
\end{figure}

In this article, we propose a graph-aligned random partition model (GARP) that exploits the flexible, but tractable, building blocks of Gibbs-type priors to build a random partition aligned on a graph.
The desired interpretation of clusters as vertices and edges in a graph naturally gives rise to dependent priors on cluster-specific parameters.
In the motivating example with single-cell RNA-seq data, vertex-clusters represent homogeneous cell subpopulations and edge-clusters correspond to cells that are transitioning between those. 
See Figure \ref{fig: mice data} for a scatter plot of single-cell RNA data in a two-dimensional space that captures most of the recorded genetic expressions of mice stem cell data.

The remainder of the article is as follows. 
In Section \ref{sec: model} we introduce a model for graph-aligned probabilistic clustering.
In Section \ref{sec: comp Gibbs} we introduce special examples.
In Sections \ref{sec: understanding}, \ref{sec: fin exc} and \ref{sec: posterior inference} we study a useful approximation, implied homogeneity assumptions, and identifiability of vertices versus edges.
Section \ref{sec: application} applies the model to single-cell RNA-seq data of mice stem cells and Section \ref{sec: discussion} concludes with final comments.
Substantive additional details, including proofs, validations on simulated data, a characterization in terms of discrete probabilities, a discussion of hyperparameter choices, and details on the strategy to obtain point estimates from posterior samples are available as an online supplement. 
The code is available at 
\if1\blind
\url{https://github.com/GiovanniRebaudo/GARP}.
\fi
\if0\blind
\url{https://github.com/BlindedStat/GARP}.
\fi

\section{Graph-Aligned Random Partition Model}
\label{sec: model}
We introduce a graph-aligned random partition model (GARP) for $\yb = \{\yb_{i}: i=1,\ldots, N\}$, $\yb_{i} \in \Re^d$.
The two main features of the model are a two-level random partition structure that assigns observations into vertex-clusters and edge-clusters, and a mixture of normal sampling models with cluster-specific parameters that reflect this split into vertex and edge-clusters.
That is, the mixture of normal models is set up such that observations in vertex-clusters form homogeneous subsets in Euclidean space, and observations in edge-clusters are located between the adjacent vertices. 
We characterize the model in three different representations that are minor variations of representations that are traditionally used for infinitely exchangeable random partition models \citep{pitman1996some}, including (1) the probability mass function (pmf) of the graph-aligned random partition via the introduction of exchangeable partition probability functions (EPPF); (2) a composition of P\'olya urn schemes, i.e., predictive probability functions, using a generalized CRP (gCRP); and (3) the configuration of ties that is implied by sampling from a composition of discrete random probability measures, similar to the construction of species sampling processes (SSP).
See \cite{pitman1996some} and \cite{lee2013defining} for details on these three characterizations for infinitely exchangeable random partitions (without alignment on a graph).

\subsection{A Gaussian Mixture over Vertices and Edges}
\label{sec: vert edge}
We start the model construction with a sampling model given the latent graph-aligned partition.
We need some notation.
Let $V_{i}$ be an indicator for observation $i$ being placed into a vertex-cluster and let $Z_{i}$ denote a cluster membership indicator.
We write $\Vb=(V_{1},\ldots,V_{N})$ and $\Zb = (Z_{1},\ldots,Z_{N})$ (throughout $\xb$ denotes the collection of all previously defined elements $x_a$).
We denote with $N_{v, N} = \sum_{i=1}^{N} V_{i}$ the number of observations in vertex-clusters, and with $N_{e, N} = N-N_{v, N}$ the implied number in edge-clusters.
For notational simplicity, we drop the subscript $_{N}$ when implied by the context. 
If $i$ belongs to a vertex (i.e., $V_{i}=1$), then $Z_{i}\in [K_{v}] \equiv \{1,\ldots,K_{v}\}$, where $K_{v}$ is the random number of vertex-clusters.
If $i$ belongs to an edge (i.e., $V_{i}=0$), then $Z_{i} = (\kv,\kvp)$, with $\kv < \kvp$ indicating the adjacent vertex-clusters.
Let $K_{e}$ denote the number of edge-clusters.
Clearly, an edge must connect two vertices, implying $K_{e} \le \frac{K_{v}(K_{v}-1)}{2} \equiv M_{e}$.
Finally, let $\Zv=(Z_{i}: V_{i}=1)$ and $\Ze=(Z_{i}: V_{i}=0)$ denote the set of cluster membership indicators for vertices and edges, respectively. 

Given a graph-aligned random partition, we assume normal sampling 
\begin{equation} \label{eq: likelihood}
    \yb_{i} \mid Z_{i}, \mubs, \Sigbs \indsim \Norm(\yb_{i} \mid \mubs_{Z_{i}}, \Sigbs_{Z_{i}}), \quad (i=1,\ldots,N),
\end{equation}
keeping in mind that $Z_{i}=k$ for $V_{i}=1$ and $Z_{i}=(k,\kvp)$ for $V_{i}=0$.
The cluster-specific parameters are defined as follows.
For the vertex-parameters $\thbs_{k}=(\mubs_{k},\Sigbs_{k})$ we assume (conditionally) conjugate normal-inverse Wishart priors
\begin{equation} \label{eq: prior NIW}
    \thb_{k}^\ast \mid K_{v} \iidsim \NIW(\mub_{0}, \lambda_{0}, \kappa_{0},\Sigb_{0}), \quad (k=1,\ldots,K_{v}).
\end{equation}
For edge-clusters, cluster-specific parameters $\thbs_{\kv,\kvp} = (\mubs_{\kv,\kvp}, \Sigbs_{\kv,\kvp})$ are defined as functions of the adjacent vertex-clusters,
\begin{equation} \label{eq: prior edge atom}
    \mubs_{\kv,\kvp} = \frac{\mubs_{\kv}+\mubs_{\kvp}}{2},~~	\Sigbs_{\kv,\kvp} = f(\mubs_{\kv},\mubs_{\kvp},r_{0},r_{1}).
\end{equation}
Here $f$ is such that the $\alpha\%$-level contour of the $N(\mubs_{\kv,\kvp}, \Sigbs_{\kv,\kvp})$ density is stretched around the line $L_{\kv,\kvp}$ connecting $\mubs_{\kv}$ and $\mubs_{\kvp}$, the Gaussian component projected onto $L_{\kv,\kvp}$ has standard deviation $r_{0}\, ||\mubs_{k}-\mub_{\kvp}^\ast||$, and the projection onto the orthogonal complement $L_{\kv,\kvp}^\perp$ are $d-1$ independent Gaussian distributions with variances $r_{1}^{2}$. 
Figure \ref{fig: Gauss edge contour supp} in Section \ref{sec: mv edge Gauss supp} of the supplemental materials shows the contour plot of an edge-cluster in $\Re^{2}$. 
See the same section and Section \ref{sec: hyper supp} of the supplementary materials for more discussion of $\Sigbs_{\kv,\kvp}$, and comments on the choice of hyperparameters $r_{0}, r_{1}$.

\subsection{Graph-Aligned Random Partition (GARP)} \label{sec: GARP}
We introduce a flexible graph-aligned random partition model. 
In words, we first label each item as belonging to a vertex or edge cluster (with probability $p_{v}$ and $(1-p_{v})$, respectively), then use a Gibbs-type prior to cluster items associated with vertices, and a Dirichlet-multinomial prior to place those associated with edges into one of the $M_{e}$ possible edges, respectively. 
Let $(n_{1},\ldots, n_{K_{v}})$ denote the cardinalities of the vertex-clusters, i.e., $n_{k} = \sum_{i} \Ind(\{V_{i}=1\} \cap \{Z_{i}=k\})$, and similarly let $n_{\kv,\kvp}=\sum_{i} \Ind(\{V_{i}=0\} \cap \{ Z_{i}=(k,\kvp)\})$ denote the sizes of the implied edge-clusters, with $n_{\kv,\kvp}=0$ indicating the lack of an edge between $k,\kvp$.
We define a graph-aligned random partition model via the pmf of $\Vb,\Zb$ 
\begin{align} \label{eq: pr V Z}
    \begin{split}
        \gn(\Vb,\Zb) &\propto p_{v}^{N_{v}}\, \EPPF_{K_{v}}^{(N_{v})}(n_{1},\ldots,n_{K_{v}}\mid \alpha, \sigma)/K_{v}! \\
	&(1-p_{v})^{N_{e}} \DM_{M_{e}}^{(N_{e})}((n_{\kv,\kvp})_{k<\kvp} \mid	\beta/M_{e})\,		\Ind(\underbrace{\{N_{e}=0\} \cup \{M_{e}>0\}}_{\Tr}), 
    \end{split}
\end{align}
where $\EPPF(\cdot \mid \alpha, \sigma)$ denotes the EPPF of a Gibbs-type prior, $\DM$ is the marginal likelihood of an $M_{e}$-symmetric Dirichlet-multinomial model (for categorical realizations, and defining $\DM_{\cdot}^{(0)}(\cdot) = \DM_{0}^{(\cdot)}(\cdot) \equiv 1$) and $\Ind(\{N_{e}=0\} \cup \{M_{e}>0\})$ is an indicator that represents the constraint that edges can only be assigned if there are at least 2 vertices ($M_{e}>0$, that is, $K_{v}>1$), or no units are assigned to edges ($N_e=0$).
We will use $\Tr$ to refer to this truncation event.
In particular, when $K_{v}=1$ (and therefore $M_{e}=0$) \eqref{eq: pr V Z} reduces to $\gn(\Vb,\Zb) \propto p_{v}^{N}\, \EPPF_{1}^{(N)}(N\mid \alpha, \sigma)$ with $V_{i}=Z_{i}=1$, for all $i$, and $\gn(\Vb,\Zb) =0$ for any other configuration $(\Vb,\Zb)$, e.g., any configuration with $N_{e}>0$ (i.e., $\Tr^{c}$).

An EPPF characterizes the distribution of an exchangeable partition \citep{pitman1996some}, with $\EPPF_{K_{v}}^{(N_{v})}(n_{1},\ldots,n_{K_{v}})$ being the probability of observing a particular (unordered) partition of $N_{v}$ observations into $K_{v}$ subsets of cardinalities $\{n_{1},\ldots,n_{K_{v}}\}$. 
Since an EPPF refers to unordered partitions we include the additional denominator $K_{v}!$ for the ordered $\Zb$.
See Section \ref{sec: understanding} for more discussions of the homogeneity assumptions implied by our model.
We specify the EPPF as a Gibbs-type prior,
\begin{align} \label{eq: EPPF Gibbs}
    \EPPF_{K_{v}}^{(N_{v})}(n_{1},\ldots,n_{K_{v}}\mid \alpha, \sigma) = W_{N_{v},K_{v}} \prod_{\kv=1}^{K_{v}} (1-\sigma)_{n_{\kv}-1},
\end{align}
where $(x)_{n}=x(x+1)\ldots(x+n-1)$ represents the ascending factorial, $\sigma<1$ is a discount parameter and the set of non-negative weights $\{W_{n,k}: 1 \le k \le n \}$ satisfies the recursive equation $W_{n,k} = (n - \sigma k) W_{n+1,k} + W_{n+1,k+1}$.
The parameter $\alpha$ in the conditioning set is used to define $W_{n,k}$ for some of the upcoming examples.
In a second step, the observations assigned to edges are (ordered) clustered using a DM distribution. 
\begin{multline}
    \gn((Z_{i}:~ V_{i}=0)\mid \Vb, K_{v}) = \DM_{M_{e}}^{(N_{e})}((n_{\kv,\kvp})_{\kv < k_{v}^{\prime}} \mid \beta/M_{e}) \\
    =\frac{\Gamma(\beta) }{\Gamma(N_{e} + \beta/M_{e})}	\prod_{(k,\kvp): k<\kvp \le K_{v}} \frac{\Gamma(n_{\kv,\kvp} + \beta/M_{e})}{\Gamma(\beta/M_{e})}.
    \label{eq: DM marginal}
\end{multline}
Model \eqref{eq: pr V Z} is a hierarchical constrained composition of a Gibbs-type prior and a symmetric-DM with hyperparameter $\beta/M_{e}$.
As we shall show, the model preserves most of the analytical and computational tractability of the simpler building blocks.

\subsection{Generalized Chinese Restaurant Process} \label{sec: gCRP}
In an alternative characterization of \eqref{eq: pr V Z}, the model can be defined as a truncated version of a composition of gCRP.
We denote the latter, that is, the model before the truncation, as $\gnt$ and refer to it as the \emph{relaxed model}.
\begin{equation} \label{eq: trunc urn}
	\gn(\Vb,\Zb) \propto \tildegn(\Vb,\Zb)\Ind(\Tr).
\end{equation}
Recall that $\Tr= \{N_{e}=0\} \cup \{M_{e}>0\}$ is the truncation. 
In Section \ref{sec: understanding} we show that $\gnt$ assigns high probability to $\Tr$, going to $1$ with $n \to \infty$ for most Gibbs-type priors.

The relaxed model $\tildegn(\Vb,\Zb)$ is a hierarchical composition of tractable generalized P\'olya urn schemes, starting with the assignments to vertices or edges 
\begin{equation} \label{eq: prior V_i}
    V_{i} \iidsim \text{Bern}(p_{v}), \quad (i=1,\ldots,N).
\end{equation}
Next, we sample cluster membership indicators $\Zv = (Z_{i}:\; V_{i}=1)$ for the vertex-clusters from the gCRP associated with Gibbs-type prior, i.e., $\Zv \mid \Vb \sim \text{gCRP}(\alpha, \sigma)$, with the gCRP implied by $\gnt$ given as 
\begin{equation} \label{eq: Gibbs urn}
  \gnt \{Z_{i}=\kv \mid \Zb\ii, \Vb\ii, V_{i}=1\}
  = \begin{cases}
		\frac{W_{N_{v},K_{v}\ii}}{W_{N_{v}-1,K_{v}\ii}} (n_{\kv}\ii-\sigma) & \kv \in [K_{v}\ii]
		\vspace*{0.2cm}\\
		\frac{W_{N_{v},K_{v}\ii+1}}{W_{N_{v}-1,K_{v}\ii}} & \kv = K_{v}\ii +1.
	\end{cases}
\end{equation}
Throughout $\mathbf{x}\ii$ identifies a quantity after removing the element $i$ from $\mathbf{x}$. 
Moreover, we use the following notation in the manuscript: given a probability measure $P$ we denote by $P\{E\}$ the probability measure evaluated in a set $E$ and by $P(a)$ the corresponding probability density function (pdf) or pmf evaluated in a point $a$.
See Section \ref{sec: comp Gibbs} for examples of different gCRP and implied prior assumptions on the number of vertices.

Finally, the cluster membership indicators $\Ze$ for the observations in edges follow the P\'olya urn scheme induced by a DM distribution
\begin{equation} \label{eq: DM urn}
	\gnt \{ Z_{i}=(\kv,\kvp)\mid V_{i}=0, \Zb\ii, \Tr\}	\propto n_{\kv,\kvp}\ii + \beta/M_{e},
\end{equation}
with $\kvp<k \le K_{v}$. 
Here, $\beta/M_{e}$ favors sparsity as the dimension of the graph increases.
Note that \eqref{eq: prior V_i} might generate $N_{e}>0$, even when \eqref{eq: Gibbs urn} implies $M_{e}=0$. 
For this case we define for completeness $\tildegn \{Z_{i}=(1,2)\mid V_{i}=0, \Zb\ii, \Tr^{c}\} \equiv 1$ (without implications for $\gn$, due to the inclusion of the truncation to $\Tr$ in \eqref{eq: trunc urn}). 

The aforementioned composition of urn schemes characterizes the GARP \eqref{eq: pr V Z}:
\begin{Prop} \label{prop: GARP vs urn}
    The random partition structure of the {\rm{GARP}} model \eqref{eq: pr V Z} can be characterized as the truncated composition of {\rm{gCRP}} defined in \eqref{eq: trunc urn}, \eqref{eq: prior V_i}, \eqref{eq: Gibbs urn} and \eqref{eq: DM urn}. 
\end{Prop}
We rely on this representation to derive an MCMC algorithm that generalizes the marginal MCMC algorithms for DP mixture models and Gibbs-type priors \citep{neal2000markov,deblasi2015gibbs,miller2018mixture}. 
Moreover, as we shall see, the probability of the truncation event $\Tr$ is high and rapidly goes to $1$ in most cases. 

\paragraph{Composition of Discrete Random Probabilities.} 
Finally, in Section \ref{sec: discr supp} of the supplementary materials we derive a third characterization of the proposed GARP. 
We define $\gnt$ as a graph-aligned random partition (with unique atoms) implied by the ties under conditionally i.i.d.\ sampling of $\thb_{i}$. 
Such a characterization will be used in a lemma to prove Theorem \ref{th: limit de Fin Dir} and can be used to connect with existing BNP literature to derive a conditional Gibbs sampler.

\section{Specific Model Choices\label{sec: comp Gibbs}} 
Conditioning on the vertex assignments $\Vb$, under the relaxed model $\gnt$ the distribution of the clustering indicators $\Zbv$ is given by the EPPF of a Gibbs-type prior \citep{gnedin2006exchangeable, deblasi2015gibbs}. 
We introduce four specific choices, stating the $\EPPF_{K_{v}}^{(N_{v})}(n_{1},\ldots,n_{K_{v}})$ for partitioning $N_{v}$ observations into $K_{v}$ vertices.
Table \ref{tab: PPF} shows the corresponding expressions for $\gnt\{Z_{i}=\kv \mid V_{i}=1, \Zbv\ii, \Vb\ii\}$ in  the gCRP of \eqref{eq: Gibbs urn}, and the weights and atoms for $P_{v}=\sum_{m=1}^{M_{v}} \pi_{m} \delta_{\tilde{\thb}_{m}}$ in (S.1) of the supplementary materials. 
Throughout, the prior for cluster-specific parameters remains the NIW in \eqref{eq: prior NIW}. 

\begin{table}[ht]
    \caption{$\gnt\{Z_{i}=\kv \mid \ldots \}$ in the gCRP \eqref{eq: Gibbs urn}, and weights $(\pi_{m})_{m=1}^{M_{v}}$ for $P_{v}=\sum_{m=1}^{M_{v}} \pi_{m} \delta_{\tilde{\thb}_{m}}$ in (S.1).
       See the text for the definition of examples 1 through 4.}
	\begin{center}
		{\renewcommand{\arraystretch}{1.5}
			\begin{tabular}{c|l:l|l|l}
				& \multicolumn{2}{l|}{\hspace{.5cm}
					$\gnt\{Z_{i}=\kv \mid V_{i}=1,\Zbv\ii,\Vb\ii\}\propto$} 
				& $P(\pi_{1},\pi_{2}, \ldots \mid M_{v}) $ & $p(M_{v}=m)$\\
				Ex. & $\kv \in \Zbv\ii$ & $\kv = K_{v}\ii +1$ & \\
				\hline
				1 
				& $n_{\kv}\ii + \rho $
				& $
				\rho (M_{v}-K_{v}\ii)\;^{(a)}$ 
				& $ \Dir(\rho,\ldots,\rho)$ 
				& fixed $M_{v} \in \mathbb{N}$ \\
				2 
				& $(n_{\kv}\ii+1)\; \times$ 
				& $(K_{v}\ii)^{2}-K_{v}\ii \gamma$
				& $\Dir(1,\ldots,1)$ 
				& $\frac{\gamma (1-\gamma)_{m-1}}{m!}$\\
				& $~~ (N_{v}\ii-K_{v}\ii+\gamma)$ 
				& 
				& 
				& \\
				3 
				& $ n_{\kv}\ii$ 
				& $\alpha$ 
				& $\GEM(\alpha)\; ^{(b)}$
				& $M_{v}=\infty$ \\
				4 
				& $n_{\kv}\ii-\sigma$
				& $ \alpha +K_{v}\ii \sigma $ 
				& $\GEM(\alpha,\sigma)\; ^{(b)}$ & $M_{v}=\infty$
		\end{tabular} }
	\end{center}
	$^{(a)}$ subject to $K_{v}\ii<M_{v}$.\\
        $^{(b)}$ $\GEM$ stands for the distribution of probability weights after Griffiths, Engen, and McCloskey \citep{ewens1990population}, using the 1-parameter version defined there and the related 2-parameters extension.
	\label{tab: PPF}
\end{table}

\begin{Exa}[$M_{v}$-dimensional symmetric Dirichlet]
    \label{ex: sym dir}
    If prior information on an upper bound $M_{v}$ on the number of vertices is available we can proceed with a finite-dimensional symmetric Dirichlet prior \citep{green2001modelling}. 
     \begin{equation}
         \label{eq: EPPF Dir}
         \EPPF_{K_{v}}^{(N_{v})}(n_{1},\ldots,n_{K_{v}}) =  \frac{M_{v}!}{(M_{v}-K_{v})!} \frac{\Gamma( \rho  \, M_{v}) }{\Gamma(N_{v} +  \rho \, M_{v}) \Gamma(\rho )^{K_{v}}} \prod_{\kv=1}^{K_{v}} \Gamma(n_{\kv} + \rho ).
     \end{equation}
     \vspace*{-0.5cm}
 \end{Exa}
Allowing for unknown $M_{v}$ the model becomes a mixture of symmetric Dirichlet model, that is, a mixture of finite mixtures (MFM).   
MFMs can be particularly interesting for allowing consistent estimation of any finite number of clusters \citep{nobile1994bayesian,miller2018mixture}.
MFMs are a special case of Gibbs-type priors.
A relevant example is the {\em Gnedin process}.
\begin{Exa}[Gnedin process, with $\sigma=-1$]
    \label{ex: GP}
    Under the Gnedin prior with parameter $\gamma \in (0,1)$ the $\EPPF_{K_{v}}^{(N_{v})}$ in \eqref{eq: pr V Z} becomes
    \begin{equation*}
      \EPPF_{K_{v}}^{(N_{v})}(n_{1},\ldots,n_{K_{v}}) =
      \sum_{m=1}^{\infty} \EPPF_{K_{v}}^{(N_{v})}(n_{1},\ldots,n_{K_{v}}\mid M_{v}=m)\, p(M_{v}=m),
    \end{equation*}
    where $\EPPF_{K_{v}}^{(N_{v})}(n_{1},\ldots,n_{K_{v}}\mid M_{v}=m)$ is the {\rm{EPPF}} of the $M_{v}$-symmetric Dirichlet prior in \eqref{eq: EPPF Dir}, with $\rho=1$ and
$	
 p(M_{v}=m) = \frac{\gamma (1-\gamma)_{m-1}}{m!}.
$
\end{Exa}
The gCRP for the Gnedin process allows tractable analytical results and efficient algorithms.
Moreover, the Gnedin process entails a distribution on the number of components $M_{v}$ that has the mode at $1$, a heavy tail, and infinite expectation \citep{gnedin2010species}. 
Therefore, the implied MFM favors a small number of vertices, while also being robust due to the heavy tail distribution of $M_{v}$.

Note that one can use $M_{v}=\infty$ to let the number of vertices (i.e., $K_{v}$) grow to infinity with $N_{v}$. 
Examples are the DP which entails a logarithmic growth of the number of vertices and the PYP which entails a polynomial growth of the number of vertices.
\begin{Exa}[DP]
    \label{ex: DP}
    Under the DP prior with parameter $\alpha>0$ the $ \EPPF_{K_{v}}^{(N_{v})}$ in \eqref{eq: pr V Z} becomes 
    \begin{equation*}
      \EPPF_{K_{v}}^{(N_{v})}(n_{1},\ldots,n_{K_{v}}) =
      \frac{\alpha^{K_{v}}\Gamma(\alpha)}{\Gamma(\alpha+N_{v})} \prod_{\kv=1}^{K_{v}} (n_{\kv}-1)!
    \end{equation*}
\end{Exa}
\begin{Exa}[PYP]
    \label{ex: PYP}
    Under a PYP prior with parameters $\sigma \in [0,1)$ and $\alpha>0$ the $ \EPPF_{K_{v}}^{(N_{v})}$ in \eqref{eq: pr V Z} becomes
    \begin{equation*}
      \EPPF_{K_{v}}^{(N_{v})}(n_{1},\ldots,n_{K_{v}}) = \frac{\Gamma(\alpha+1) \prod_{\kv=1}^{K_{v}-1} (\alpha+\kv \sigma ) }{\Gamma(\alpha+N_{v})} \prod_{\kv=1}^{K_{v}} (1-\sigma )_{n_{\kv}-1}.
    \end{equation*}
\end{Exa}
With $\sigma=0$ the PYP reduces to the DP.
Other popular sub-classes of Gibbs-type priors include the NGPP \citep{lijoi2007controlling}, the NIG \citep{lijoi2005hierarchical,lijoi2007bayesian}, and the MFM \citep{nobile2007bayesian,miller2018mixture}.
See \cite{deblasi2015gibbs} for a comprehensive review of Gibbs-type priors.

Finally, we note that here we focus on prior elicitation of the Gibbs-type random partition that controls the vertex-clusters and the number of vertices (i.e., $K_{v} \le \min(M_{v}, N_{v}) \le \min(M_{v}, N)$). 
Given $K_{v}$ the possible number of edges is finite. 
The only Gibbs-type prior with a finite fixed number of components $M_{e}$ is the symmetric Dirichlet \citep[see e.g,][]{deblasi2015gibbs}, that is the $\text{DM}_{M_{e}}$ in \eqref{eq: pr V Z}. 
Although the preceding discussion focuses on the Gibbs-type partition that controls the vertices assignment, it entails (thanks to the hierarchical definition e.g., in Section \ref{sec: gCRP}) similar flexibility in the joint prior elicitation of the vertices assignments. 

\section{Goodness of the Approximation} 
\label{sec: understanding}
We discuss properties of the approximation of the GARP model in \eqref{eq: pr V Z} by the {relaxed model} $\tildegn$, and why it is a good approximation of $\gn$, justifying the prior elicitation of $\gn$ via $\tildegn$.
Importantly, the results allow us to effectively sample from the GARP via rejection sampling, using proposals from $\tildegn$. 
\begin{Prop} \label{prop: prob trunc}
    The probability of the truncation event $\Tr$  under the relaxed model	is
    \begin{equation}
        \gnt\{\Tr\} = p_{v}^{N} + \sum_{n_{v}=2}^{N-1} \binom{N}{n_{v}} p_{v}^{n_{v}} (1-p_{v})^{(N-n_{v})} \big[1-(1-\sigma)_{n_{v}-1} W_{n_{v},1} \big].
        \label{eq: prop prob trunc}
    \end{equation}
\end{Prop}
Here $p_{v}^{N}=\gnt\{N_{v} = N\}$, and $(1-\sigma)_{n_{v}-1} W_{n_{v},1}$ in the second term arises from \eqref{eq: EPPF Gibbs} as the probability given $\{N_{v}=n_{v}\}$ of having a single vertex, i.e., $\gnt \{ K_{v}=1 \mid N_{v}=n_{v}\}= \EPPF_{1}^{(n_{v})}(n_{v})$.
For the Gibbs-type priors in the following examples, the latter reduces to simple analytical expressions.

In the upcoming discussion, we introduce several closely related distributions. 
To avoid confusion we provide a summary and list of defined distributions in Table \ref{table: dist supp} in the supplementary materials. 
Let $\gtn{N}$ denote the law of $V_{i}$, $i=1,\ldots,N$ and $\Zv= (Z_{i}:\; i\in[N], V_{i}=1)$ under the relaxed model. 
More precisely, $\gtn{N}$ is the joint law of the random variables $(T_{1},\ldots,T_{N})$, where $T_{i}=V_{i}$ if $V_{i}=0$ and $T_{i}= (V_{i},Z_{i})$ if $V_{i}=1$.
Let $\gt$ denote the law of the stochastic process with Kolmogorov consistent finite dimensional $(\gtn{N})_{N \in \N}$.
Such a process exists due to the i.i.d.\ nature of $V_{i}$ and the exchangeable nature of the Gibbs-type prior that defines $\Zbv$ given $\Vb$.
We therefore have by the strong law of large numbers $\lim_{N \to \infty} N_{v}/N = p_{v}$, $\gt$-a.s.
Also, note that the truncation event $\Tr$ is a function of $(\Vb,\Zbv)$ (thus $\mathbf{T}$) only, allowing us to evaluate $\gnt\{\Tr\}$ in \eqref{eq: prop prob trunc} as probabilities under $\gt$.  

We are now ready to analyze \eqref{eq: prop prob trunc}.
First, note that $\Tr^{c}$ can be decomposed as $\Tr^{c}= \left(\{K_{v}=1\} \cap \{N_{v} \ne N\}\right) \cup \{N_{v}=0\}$ and therefore
\begin{equation}
    \gt\{\Tr^{c} \} = \gt\{K_{v}=1\} - p_{v}^{N} \gt\{K_{v}=1 \mid N_{v}=N\} + (1-p_{v})^{N},
    \label{eq: prob E_N^c}
\end{equation}
with the last term corresponding to $\gt\{N_{v}=0\}$ and the sum of the first two terms corresponding to $\gt\left\{\{K_{v}=1\} \cap \{N_{v} \ne N\}\right\}$. 
Note that $(\tildegn{\{K_{v}=1\}})_{N \in N}$ and $(\gnt\{K_{v} = 1\mid N_{v}=n_{v}\})_{n_{v} \in\N}$ (well defined for any $N = f(n) \ge n$) are non-increasing sequences of elements in $[0,1]$.
This is the case since they can be seen as the probability $\gt$ of non-increasing sequences of events.
The two sequences are thus convergent.

For any $p_{v} \in (0,1)$, $(\gt\{\Tr^{c} \})_{N \in \N}$ in \eqref{eq: prob E_N^c} has limit equal to $\lim_{N \rightarrow \infty} \tildegn{\{K_{v}=1\}}$ (since $p_{V}^N$ and $(1-p_{v})^{N}$ go to $0$).
Let then $g^{\infty}=\lim_{N \rightarrow \infty} \tildegn{\{K_{v}=1\}}$, and let $g_{v}^{\infty} = \lim_{n_{v} \rightarrow \infty} \gnt\{K_{v} = 1\mid N_{v}=n_{v}\}$.
Since $K_{v}$ depends on $Z_1, \ldots, Z_N$ only indirectly through the $N_{v}$ units allocated in $\textbf{Z}_v$ and $N_{v}/N \rightarrow p_v$ a.s.\ (see the proof of Theorem 1 for more discussion), the two limits are equal, i.e., $g^{\infty} =g_{v}^{\infty}$. 
We shall show that they equal $0$ for several Gibbs-type priors, implying that the GARP will go to the relaxed model, that is, $\gnt\{\Tr\} \to 1$ as $N \to \infty$. 
Table \ref{tab: g^inf} summarizes the results for the earlier four examples. 
We use $n_{v} \le N$ and for any sequences $a_{n}$ and $b_{n}$, we write $a_{n} \asymp b_{n}$ if and only 
$\lim_{n} a_{n}/b_{n} = 1$.

\begin{table}[ht]
    \caption{$\gnt\{K_{v} = 1\mid N_{v}=n_{v}\}$, limit $g_{v}^{\infty}$ and asymptotic rate as $n_{v} \to \infty$ for Examples \ref{ex: sym dir} ($M_{v}$-dimensional symmetric DM, with $M_{v}>1$), \ref{ex: GP} (Gnedin), \ref{ex: DP} (DP) and \ref{ex: PYP} (PYP).
    }
    \begin{center}
	{\renewcommand{\arraystretch}{1.4}
		\begin{tabular}{l|l:l:l}
		      & \multicolumn{3}{l}{\hskip 1cm $g_{n_{v}} \equiv \gnt\{K_{v} = 1\mid N_{v}=n_{v}\}$}\\
		Ex. & $g_{n_{v}}= $ & $ g_{n_{v}} \asymp$
				& $g_{v}^{\infty} \equiv \lim_{n_{v}\to\infty} g_{n_{v}}$ \\
				\hline
            1   & $ \frac{(\rho)_{n_{v}}}{(\rho M_{v})_{n_{v}}} \, M_{v}$
                & $ \frac{\Gamma(\rho M_{v}) M_{v}}{\Gamma(\rho)} n_{v}^{\rho(1-M_{v})}$
				& $0$ \\
		2   & $\frac{\gamma n_{v} }{\gamma+n_{v} -1}$
				& $\gamma$
				& $\gamma \in (0,1)$\\
            3   & $\frac{\Gamma(\alpha+1) (n_{v}-1)!}{\Gamma(\alpha+n_{v})}$ 
				& $\Gamma(\alpha+1) n_{v}^{-\alpha}$
				& 0 \\
            4   & $\frac{(1-\sigma)_{n_{v}-1}}{(\alpha+1)_{n_{v}-1}}$
                & $\frac{\Gamma(\alpha+1)}{\Gamma(1-\sigma)} n_{v}^{-(\alpha+\sigma)}$ 
				& 0
		\end{tabular} }
	\end{center}
    \label{tab: g^inf}
\end{table}

\begin{Thm} \label{th: prob const}
    Under the relaxed model $\gnt$ we have $g^{\infty} = g_{v}^{\infty}=\lim_{N \rightarrow \infty} \gnt\{ \Tr^{c}\}$ with $g^{\infty}= 0$ under the symmetric Dirichlet, the DP, the PYP, and $g^{\infty} = \gamma \in (0,1)$ under the Gnedin process.
    The asymptotic rates of $g_{n_{v}}$ are given in the second column of Table \ref{tab: g^inf}.
\end{Thm}

Theorem \ref{th: prob const} and \eqref{eq: trunc urn} show that performing prior elicitation and posterior simulation based on the (analytically and computationally) simpler relaxed model $\tildegn$ becomes practically attractive.
Table \ref{tab: g^inf} also provides the rate at which $\tildegn(\Tr^{c})$ (where the two models differ) converges.
For instance, when $\tildegn(\Tr^{c}) \approx 0$ (in Theorem \ref{th: prob const}), it is immediate to consider $p_{v}$ as the prior proportion of observations assigned to vertex clusters under $\tildegn$ for any sample size $N$.
Another important consequence of Theorem \ref{th: prob const} and \eqref{eq: trunc urn} is that we can effectively sample from the prior GARP model with an acceptance-rejection method that proposes a realization from the simple relaxed model $\tildegn$ having theoretical guarantees that the acceptance probability is around $1$ in most of the cases. 
Also with the convergence of $\tildegn(\Tr^{c})$ to $\gamma>0$ under the Gnedin process, the approximation remains attractive, as rejection sampling remains practically feasible with known acceptance probability $\tildegn(\Tr)$ going to $1-\gamma$ (instead of 1, under the other models), where $\gamma$ is a hyperparameter that we can control. 

Finally, in most examples, the relaxed model $\gnt$ approaches the GARP $\gn$ as the sample $N$ increases in an even stronger way.  
\begin{Thm} \label{th: const eventually}
  Under $\tildegn$  with symmetric Dirichlet, DP or PYP ($\sigma \ge 0$) in \eqref{eq: pr V Z} 
  \begin{equation} \label{eq: eventually E_n} 
	\gt\{ \Tr \text{ eventually} \} =1.
    \end{equation}
    Thus, for any $k \in \N$ and any possible set of points $a_{k}=(\mathbf{v}_{1:N+k},\mathbf{z}_{1:N+k})$ 
    \begin{equation} \label{eq: fin Dir tilde pred}
        \widetilde{G}_{VZ}\left\{\big\{ {G^{(N+k)}}( a_{k} \mid \mathbf{V}_{1:N},\mathbf{Z}_{v,N})  = \widetilde{G^{(N+k)}}(a_{k} \mid \mathbf{V}_{1:N},\mathbf{Z}_{v,N})\big\} \text{ eventually}\right\}=1. 
    \end{equation}
    Under $\tildegn$ with the Gnedin process we have $\gt\big\{ \Tr \cup \{M_{v}=1\} \text{ eventually} \big\} =1$ and $\widetilde{G}_{VZ}\left\{ \big\{G^{(N+k)}( a_{k}\mid \mathbf{V}_{1:N},\mathbf{Z}_{v,N}) = \widetilde{G^{(N+k)}}(a_{k} \mid \mathbf{V}_{1:N}, \mathbf{Z}_{v,N}) \big\} \cup \{M_{v}=1\} \text{ eventually}	\right\}=1$.
\end{Thm}
In words, almost surely either the predictive pmf under the GARP and the relaxed will eventually coincide or (under $\tildegn$ with the Gnedin process) there is only one possible vertex-cluster for any $N\in \mathbb{N}$.  
The latter has a positive probability $\gt\{M_{v}=1\} = \gamma \in (0,1)$ for the Gnedin process.

\section{Finite Exchangeability and Projectivity} 
\label{sec: fin exc}
Under the GARP the distribution of the sample is (finitely) exchangeable, that is the marginal law of $(\yb_{i})_{i=1}^{N}$ from \eqref{eq: likelihood}--\eqref{eq: pr V Z} is invariant with respect to permutations of the labels $1,\ldots, N$.
This homogeneity assumption entails that the order in which we look at the observations does not affect the prior and the inferential results, as it should. 
The same homogeneity assumption is true for the graph-aligned random partition induced by $(V_{i}, Z_{i})_{i=1}^{N}$. 
We discuss some more details of homogeneity assumptions in the model. 
We will write $\gn$ for different distributions implied by the GARP model \eqref{eq: likelihood}--\eqref{eq: pr V Z}, with the specific distribution being clear from the argument of $\gn(\cdot)$. 

\paragraph{Finite EPPF.}
Let $\Psi_{N}$ denote the random partition of observations $[N]$ defined by clustering $i$ and $j$ together if and only if $\thb_{i}=\thb_j$ (recall that $\thb_{i}=\thbs_{Z_{i}}$).
Under the GARP model $\Psi_{N}$ is an exchangeable random partition with dependent cluster-specific parameters.
We introduce the notion of finite EPPF (fEPPF) to characterize the distribution of such random partitions: $\gn\{\Psi_{N} = \{C_{1},\ldots, C_{K}\}\} = \text{fEPPF}^{(N)}_{K}(c_{1},\ldots,c_{K})$, where $(c_{1},\dots,c_{K}) = (|C_{1}|,\ldots,|C_{K}|)$ are the cluster sizes (in a given arbitrary order).
Note that $\{c_{1},\dots,c_{K}\}$ is a sufficient statistic for an exchangeable random partition.
Here $K$ denotes the number of clusters, i.e., $K=K_{v}+K_{e}$.
The fEPPF is a symmetric function of a composition of $N$ (positive integers that sum up to $N$).
The fEPPF induced by the GARP can be obtained via marginalization of the probability function \eqref{eq: pr V Z} of the graph-aligned random partition.
Several expressions can be aggregated via probabilistic invariance.
\begin{Prop} \label{prop: fEPPF}
    Under the \rm{GARP}
    \begin{multline}
        \mathrm{fEPPF}_{K}^{(N)}(|C_{1}|,\ldots,|C_{K}|) \propto \sum_{{ N_{v}=1}}^{N} \bigg\{ \binom{N}{N_{v}} p_{v}^{N_{v}} (1-p_{v})^{N-N_{v}} \\
        \sum_{{ K_{v}=1}}^{M_{v}} \bigg[ \binom{M_{e}}{K-K_{v}} \sum_{{ (n_{1},\ldots,n_{K_{v}})}} \EPPF^{(N_{v})}_{K_{v}}(n_{1},\ldots,n_{K_{v}}) \mathrm{DM}^{(N-N_{v})}_{M_{e}} ((n_{\kv,k_{v^{\prime}}})_{\kv< k_{v}^{\prime}}). \bigg] \bigg\}
	\label{eq: fEPPF}
    \end{multline}
    In the last sum, for given $(n_{1},\ldots,n_{K_{v}})$ the	cardinalities $n_{\kv,\kvp}$ of edge-clusters are implied by the remaining elements of $(|C_{1}|, \ldots, |C_{K}|)$ that are not matched with the vertex-cluster cardinalities $n_k$.
    The exact range of the sums is stated in Section \ref{sec: proof fEPPF} of the supplementary materials.
    Essentially, $\{n_{1},\ldots,n_{K_{v}}\} \cup \{n_{\kv,\kvp}:\; k<\kvp\} = \{c_{1},\ldots, c_{K}\}$. 
    Moreover, the normalization constant in \eqref{eq: fEPPF} is $1/\gnt\{\Tr\}$, which we studied in detail before.
\end{Prop}
A common stronger assumption in the literature on random partitions is that the observed data $(\yb_{i})_{i=1}^{N}$ are a subset of an infinite (thus unobservable) sequence of exchangeable random variables.
This assumption does not apply to the GARP -- see below. 
However, if the assumption applies then the exchangeable random partition of the sample can be seen as a projection of an exchangeable random partition of the natural numbers $\mathbb{N}$ to the set $[N]$.
Formally, this is equivalent to assuming:
\begin{itemize}\itemsep=0pt
    \item[(a)] each random partition $\Psi_{N}$ is exchangeable over $[N]$;
    \item[(b)] the sequence of random partitions $(\Psi_{N})_{N=1}^{\infty}$ is Kolmogorov consistent, that is, $\Psi_{n}$ is equal in distribution to the restriction of $\Psi_{N}$ to $[n]$ for any $1 \le n \le N$. 
\end{itemize}
Note that, although we stated the properties for the random partition, the same definitions hold for other sequences of random variables, such as the sample $(\yb_{i})_{i=1}^{N}$.
As done in, e.g., \cite{betancourt2022random} we refer to (a) as \emph{finite exchangeability}, (b) as \emph{projectivity}, and to their combination as \emph{infinite exchangeability}. 

\begin{Prop} \label{prop: fin exch no inf}
  The graph-aligned random partition induced by $(V_{i}, Z_{i})_{i=1}^{N}$, the sample $(\yb_{i})_{i=1}^{N}$ and the random partition $\Psi_{N}$ are finitely exchangeable but they are not a projection of infinite exchangeable processes. 
\end{Prop}

From a modeling perspective, infinite exchangeability is a natural requirement only to address prediction problems in the most general framework, i.e., prediction for an unbounded number of future observations. 
In general, it is a desirable property for mathematical convenience to ease prior elicitation (e.g., via de Finetti's representation theorem), to simplify posterior inference, and to study the properties of the model across sample sizes.
While the GARP is not infinitely exchangeable, as stated in the previous result, in some cases it turns out to be very close to infinite exchangeability, in the sense that the model is equivalent to an infinitely exchangeable model for large enough $N$, as discussed next. 
See also \cite{diaconis1980finite} for general results and probabilistic characterizations of finite exchangeability and approximate projectivity.  
The next result shows that in some cases the predictive distribution of the GARP model eventually (i.e., for a large enough sample size $N$) can be characterized as a projection of the predictive of a limiting infinitely exchangeable model, thus where projectivity holds.

We also characterize the limit via the directing measure, i.e., the law of the random probability in de Finetti's representation theorem.
See Table \ref{table: dist supp} for a recap of the notation for different distributions.
\begin{Thm} 
\label{th: limit de Fin Dir}
  Under the {\rm{GARP}} model with the {$M_{v}$-symmetric Dirichlet} (Example \ref{ex: sym dir}) in \eqref{eq: pr V Z} there exists a finite random sample size $\bar{N}$ and an infinite dimensional law $\ginf$, such that for any $N>\bar{N}$ the predictive distributions under the GARP model, are $\gt$-almost surely equal to the predictive distributions under the (Kolmogorov consistent) marginal laws $\big(G_{N}^{(\infty)}\big)_{N\in \N}$ of the infinite-dimensional law $\ginf$. 

  That is, for any possible sequence of sets of points $(a_{k})_{k\in\N}$, with $a_{k}=(\mathbf{v}_{1:N+k}, \mathbf{z}_{1:N+k})$ 
  \begin{equation}\label{eq: limit urn de Fin}
    \gt\left\{\big\{G_{N+k}^{(\infty)}( a_{k} \mid \mathbf{V}_{1:N},\mathbf{Z}_{v,N}) = G^{(N+k)}( a_{k} \mid \mathbf{V}_{1:N},\mathbf{Z}_{v,N})\; \forall\, k\big\} \text{ eventually}\right\}=1.
  \end{equation}
  Here $\ginf$ can be characterized by the following gCRP.
  Let $\Mep=M_{v}(M_{v}-1)/2$. 
  \begin{multline} \label{eq: limit urn dir}
    \ginf\{V_{i}=v, Z_{i}=z \mid \cdots, \mathbf{V}_{1:N},\mathbf{Z}_{1:N}\} \propto 
    \begin{cases}
      p_{v} \frac{n_{\kv}\ii + \gamma}{N_{v}\ii+\gamma M_{v}} &\text{if } v=1, \quad z \in [M_{v}]\\ 
      (1-p_{v}) \frac{\beta/\Mep+n_{\kv,\kvp}\ii}{\beta/\Mep+N_{v}\ii} &\text{if } v=0, \quad z = (\kv, \kvp).
    \end{cases}
  \end{multline}
  The directing measure characterizing the infinitely exchangeable random parameters that imply $\ginf$ is defined as
  \begin{equation} \label{eq: limit de Fin Dir}
    (\mub_{i}, \Sigb_{i}) \mid P \iidsim P,\quad \quad P = p_{v} \sum_{m=1}^{M_{v}} \pi_{m} \delta_{\thbt_{m}} + (1-p_{v}) \sum_{k<\kvp { \le M_{v}}} \pi_{\kv,\kvp} \delta_{\thbt_{\kv,\kvp}} 
  \end{equation}
  where $(\pi_{1},\ldots,\pi_{M_{v}}) \sim \Dir({ \rho,\ldots,\rho})$, $(\pi_{\kv,\kvp})_{k<\kvp} \sim \Dir(\beta/\Mep,\ldots,\beta/\Mep)$, and $\thbt_{m}$ and $\thbt_{\kv,\kvp}$ follow the same distributions as in \eqref{eq: prior NIW} and \eqref{eq: prior edge atom}. 

  Let $\ginf(\Vb,\Zb)$ denote the pmf of $\Vb,\Zb$ implied by \eqref{eq: limit de Fin Dir}.
  It can also be characterized by the projective pmfs for any $N\in \mathbb{N}$ (we omit the sub-index $_{N}$ for the finite projections of $\ginf$ when it is clear from the context):
  \begin{multline} \label{eq: G_inf}
    \ginf((V_{i},Z_{i})_{1:N})= p_{v}^{N_{v}} \EPPF_{M_{v}}^{(N_{v})}(n_{1},\ldots,n_{K_{v}}\mid \alpha, \sigma)/K_{v}!
    \\
    \times (1-p_{v})^{N_{e}} \DM_{\Mep}^{(N_{e})}((n_{\kv,\kvp})_{\kv< k_{v}^{\prime}} \mid \beta/ \Mep ).
  \end{multline}
\end{Thm}
\begin{Cor} \label{cor: limit MFM}
  Conditional on a given $M_{v}$, Theorem \ref{th: limit de Fin Dir} remains true also under the {\rm{GARP}} with a Gnedin process (Example 2), with $\ginf(M_{v}=m)= \gt(M_{v}=m)= \frac{\gamma (1-\gamma)_{m-1}}{m!}$.
\end{Cor}
See \ref{sec: proof supp} for the explicit statement of Corollary \ref{cor: limit MFM} and the proofs. 

Analogous results hold for any MFM. 
We state it for the special case of the Gnedin process which we introduced and discussed in Section \ref{sec: comp Gibbs}. 

Note that even if projectivity is not strictly needed to carry out inference under the GARP, approximate projectivity is still a useful property.
Without any form of approximate projectivity (i.e., coherence), inference on the partition structure for $N$ observed units would depend on whether or not an investigator plans to collect more data in the future. 
This would greatly complicate the understanding of model assumption and learning mechanisms.

\section{Posterior Inference}\label{sec: posterior inference}
Building on the earlier results we develop MCMC algorithms for posterior simulation under the GARP.
The algorithms generalize the posterior sampling scheme for the CRP under a DP mixture \citep{neal2000markov} and under Gibbs-type mixtures.
To derive tractable full conditional distributions that are easy to sample from, we exploit the representation of the GARP as a truncated composition of Gibbs-type priors derived in Section \ref{sec: gCRP}.

In this way, we can exploit the product partition form of the pmf under the relaxed model to simplify the expressions of the conditional probability in the predictive (i.e., the composition of gCRPs) and full conditional distributions.
Expressions reduce to simple ratios.

In general, without projectivity and composition of product partition EPPF, it is not possible to generalize a priori (and a posteriori) tractable P\'olya urn schemes and thus tractable marginal algorithms such as the ones in \cite{neal2000markov}.   
Projectivity allows us to evaluate conditional probabilities (of cluster membership) as ratios of the same EPPFs over different $N$. 
Under the specific product form of the EPPF for Gibbs-type priors, this ratio reduces to a simple expression \citep{deblasi2015gibbs}.
 
Specifically, the relaxed model $\tildegn$ is a hierarchical composition of Kolmogorov consistent EPPFs with product partition forms (Sections \ref{sec: GARP} and \ref{sec: comp Gibbs}) that thus induce a tractable (a priori) composition of gCRPs (Section \ref{sec: gCRP}). 
This allows us to derive the following efficient marginal sampler.
See Section \ref{sec: neal} in the supplementary materials for details.

For an explicit statement of Gibbs sampling transition probabilities, we introduce the notation $I_{i}=\Ind(\{n_{k}\ii=0\} \cap \{ \sum_{\kvp \ne k} \big(n_{\kv,\kvp}+n_{\kvp,k}\big)>0 \})$ as an indicator for violating the support of the GARP in \eqref{eq: pr V Z}.
That is, $I_{i}=1$ if removing $i$ from its current cluster removes the last unit in a vertex-cluster $k$ (for some $k$) and it leaves an edge-cluster $(k,\kvp)$ (for some $\kvp\ne k$) without adjacent vertex-cluster $k$. 

We then have the following full conditional probabilities. 
\newline
\texttt{(1)} Sample $(V_{i}, Z_{i})$ form $\gn(V_{i}, Z_{i}\mid \cdots)$.
If $I_{i}=1$ we do not move. 
Otherwise sample from $\gn\{V_{i}=v, Z_{i}=z \mid \cdots\} \propto$
\begin{multline}
    \quad
    \begin{cases}
        p_{v} \frac{W_{N_{v},K_{v}\ii}}{W_{N_{v}-1,K_{v}\ii}} (n_{\kv}\ii-\sigma) \Norm(\bm{\yb_{i}} \mid \mubs_{k},\Sigb_{\kv}^\ast)& \text{if } v=1, \quad z \in [K_{v}\ii]\\
        p_{v} \frac{W_{N_{v},K_{v}\ii+1}}{W_{N_{v}-1,K_{v}\ii}}  g_{\text{new}}(\yb_{i}) &\text{if } v=1, \quad z =K_{v}\ii+1\\
        (1-p_{v}) \frac{\beta/M_{e} +n_{\kv,\kvp}\ii}{\beta/M_{e}+N_{v}\ii} \Norm(\bm{\yb_{i}} \mid \mub_{\kv,\kvp}^\ast,\Sigb_{\kv,\kvp}^\ast) &\text{if } v=0, \quad z=(\kv, \kvp),
    \end{cases}
    \nonumber
\end{multline}
where 
\begin{equation*}
    g_{\text{new}}(\yb_{i}) = \int \Norm(\yb_{i} \mid \mub, \Sigb) \, \mathrm{d}\mbox{NIW}(\mub, \Sigb \mid {\mub}_{0}, \lambda_{0}, \kappa_{0}, {\Sigb}_{0})= \mbox{T}_{\lambda_{0}-1} \bigg(\yb_{i} \mid \mub_{0}, \frac{\kappa_{0}+1}{\kappa_{0}(\lambda_{0}-1)}\bigg)
\end{equation*}
is the pdf of a generalized Student-T distribution of degree $\lambda_{0}-1$.

\noindent
\texttt{(2)} Sample the vertices parameters $(\mubs_{k},\Sigb_{\kv}^\ast)$ from
$$
\gn(\mub_{\kv},\Sigb_{\kv} \mid \cdots) \propto \underbrace{\NIW\big(\mubs_{\kv},\Sigbs_{\kv} \mid \hat{\mub}, \hat{\nu},\hat{\kappa}, \hat{\Sigb}\big)}_{p^0(\thbs_{k})}\, \times \, \prod_{\kvp\ne k} \Norm(\yb_{i} \mid \mubs_{\kv,\kvp}, \Sigbs_{\kv,\kvp})
$$
where in the last product for $\kvp<k$ we interpret $\thbs_{\kv,\kvp}$ as $\thbs_{\kv,\kvp} \equiv \thbs_{\kvp, k}$, and $\hat{\nu} = \nu_{0} + n_{\kv}$, $\hat{\kappa}_{\kv} = \kappa_{0} + n_{\kv}$, $\hat{\mub} = \frac{\kappa_{0} \mub_{0} + n_{\kv} \bar{\yb}_{\kv} }{\hat{\kappa}_{\kv}}$ and	$\hat{\Sigb} = \Sigb_{0} + \bm{S}_{\kv} + \frac{\kappa_{0} n_{\kv}}{\hat{\kappa}_{\kv}} (\bar{\yb}_{\kv}-\mub_{0}) (\bar{\yb}_{\kv}-\mub_{0})^{\intercal}$, with $\bar{\yb}_{\kv} = \frac{\sum_{i: Z_{i} = \kv } {\yb}_{i}}{n_{\kv}}$ and $\bm{S}_{\kv} = \sum_{i: Z_{i} = \kv } (\yb_{i}-\bar{\yb}_{\kv})(\yb_{i}-\bar{\yb}_{\kv})^{\intercal}$. 

If a vertex is isolated, that is, no observations are assigned to any of the possible edges associated with the vertex, then the full conditional in \texttt{(2)} reduces to the conjugate NIW posterior distribution $p^0(\thbs_{k})$.
In general, the density of the full conditional is proportional to $p^0$ times the likelihood of the observations assigned to corresponding edges. 
An effective transition probability is a Metropolis-Hasting step exploiting $p^0$ as a proposal. 
\medskip

In step \texttt{(1)}, when we create a new vertex-cluster, i.e., if $v=1$ and $\kv =K_{v}\ii+1$, we follow up with a transition probability (2) for the new cluster parameters, that reduces to the conjugate NIW for $\thbs_{k}$.
Throughout, edge-parameters $\thbs_{\kv,\kvp}$ are always evaluated using the currently imputed adjacent vertex parameters $\thbs_{k}, \thbs_{\kvp}$.

Note that it is also possible to add an extra transition probability to update $Z_{i}$ as in (1), but leaving $V_{i}$ unchanged. 
Such transition probabilities could lead to a better mixing Markov chain and are analogous to the ones used, for example, in \cite{teh2006hierarchical} exploiting the Chinese restaurant franchise representation of the hierarchical DP.
\medskip

In principle, all posterior inference is implemented by appropriate summaries of the posterior Monte Carlo sample. 
However, how to report point estimates for a random partition or graph is not trivial.
There are several proposals in the recent literature, including \cite{wade2018bayesian}, \cite{dahl2022search}, and \cite{franzolini2024entropy}.
They are based on casting the selection of the reported summary as a decision problem. 
In Section \ref{sec: Dahl supp} of the supplemental materials, we discuss an implementation for the GARP.

Finally, like in any mixture model, posterior inference about specific clusters must consider label switching.
See, for example, \cite{green2018introduction} for a discussion.
An additional challenge that arises in the proposed model is the distinction between vertex versus edge clusters.
Consider, for example, a configuration (A) with $2$ vertices and a connecting edge, with cluster-specific parameters $(\thbs_{1}, \thbs_{2}, \thbs_{1,2} = f(\thbs_{1}, \thbs_{2}))$ (as in \eqref{eq: prior edge atom}), versus an alternative configuration (B) with $3$ vertices and $(\thbs_{1}, \thbs_{2}, \thbs_{3})$ and $\thbs_3 = f(\thbs_{1}, \thbs_{2})$.
While the sampling model \eqref{eq: likelihood} remains unchanged under (A) versus (B), we argue that the prior implements a strong preference for (the more parsimonious) model (A).

For given vertex parameters $\thbs_{1}$ and $\thbs_{2}$, the edge parameter $\thbs_{1,2}$ in (A) can assume just one value, i.e., its parameter space is the single point $f(\thbs_{1}, \thbs_{2})$ in the parameter space of the third vertex $\thbs_{3}$ in the latter model.
In other words, when we consider the joint parameter space $\bm{\Theta}_{0}$ of the atoms $\thbs_{1}, \thbs_{2}, \thbs_{1,2}$ (two vertices and one edge) it is a lower dimensional sub-space of the parameter space $\bm{\Theta}$ for the three vertices $\thbs_{1}, \thbs_{2}, \thbs_{3}$.
The NIW prior in \eqref{eq: prior NIW} on $\thbs_j$ assigns prior probability $0$ to $\bm{\Theta}_{0}$, and thus also zero posterior probability.
The issue is similar to identifiability related to the replication of terms in a standard mixture model with independent priors on cluster-specific parameters \cite{green2018introduction}.

\section{Application to Single-Cell RNA Data} \label{sec: application}
We fit the GARP model for the RNA-seq data shown in Figure \ref{fig: mice data}.
Single-cell RNA-seq experiments record cell-specific transcriptional profiles that allow us to infer, for example, cell differentiation or cancer progression.
Inference under the GARP model for the data shown in Figure \ref{fig: mice data} reconstructs transitions of stem cells into fully differentiated cells in a scRNA-seq experiment on horizontal basal cells from the adult mouse olfactory epithelium.
The original data is available on GEO in GSE95601.

The transcriptional profiles map differences in gene expressions due to the development phases of the cells.
Stem cells evolve into fully differentiated cells by gradual transcriptional changes, passing through a small number of homogeneous subpopulations of cells.
The primary inferential goal is to find these homogeneous subpopulations of cells (i.e., vertex-clusters) and understand the relationships between them aligning such subpopulations on a biologically interpretable graph.

\subsection{ScRNAseq Data and Pre-Processing}
The raw data is a count matrix with rows corresponding to cells and columns representing different genes.
Most of the counts in the matrix are zeros, usually about $90\%$ (the percentage can vary according to the scRNA-seq technology used).

The data set originally contains measurements for $28284$ genes in $849$ cells with $84\%$ zeros. 
To extract a lower dimensional signal we implemented pre-processing following the pipeline described in \cite{perraudeau2017bioconductor} and available in \texttt{Bioconductor} \citep{gentleman2004bioconductor}.
We briefly describe the pipeline.
We first discard around 100 low-quality cells and retain the $1000$ most variables genes. 
Next, we normalize the data matrix and extract 50-dimensional biomarkers from the count data, accounting for zero-inflation and over-dispersion of the scRNA-seq data via ``Zero-Inflated Negative Binomial Wanted Variation Extraction'' (ZINB-WaVE) \citep{risso2018general}. 
Finally, we reduced the dimensionality to the 2 most relevant markers via multidimensional scaling analysis.
The data matrix obtained after pre-processing is denoted by $\yb=(\yb_{i,j}:i=1,\ldots, N,j=1,2)$, where the rows represent $747$ cells, and the columns record the two final biomarkers.
The data is shown in Figure \ref{fig: mice data}.

\subsection{Results} \label{sec: results}
We implement inference under the GARP model using the Gnedin process (Example \ref{ex: GP}) to control the vertex-clustering.
We choose the Gnedin process because one of the goals is inference on $K_{v}$.
The Gnedin process is a particularly attractive Gibbs-type prior for clustering from both, a Bayesian modeling perspective as well as for its frequentist properties of the posterior distribution, as discussed in Section \ref{sec: comp Gibbs}.

\begin{figure}[ht]
    \centering
    \includegraphics[width=8cm]{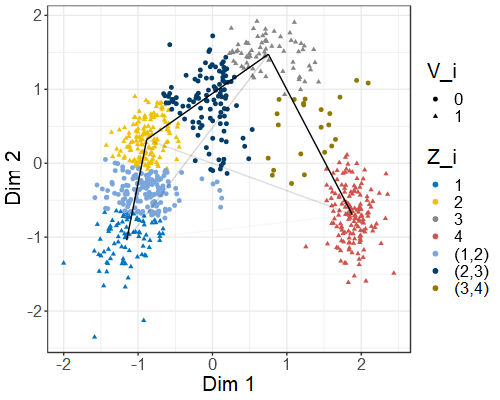}
    \includegraphics[width=8cm]{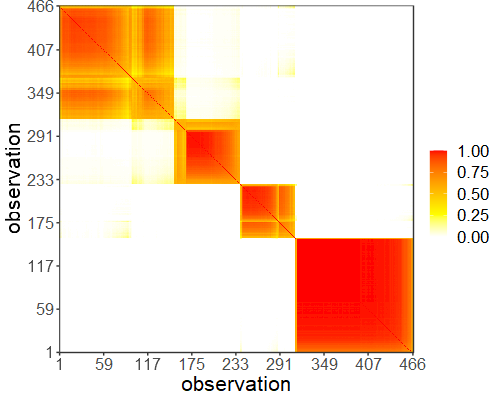}
    \\
    (a) \hspace{7.7cm} (b) \hspace{0.4cm}
    \caption{Left Panel: Scatter-plot of the scRNA data. Triangular plot symbols indicate cells assigned to vertices ($V_i=1$) while the remaining cells are assigned to edges ($V_i=0$) and are represented with a circular shape.
    Cells are colored according to the different phases (i.e., $Z_i$) in the point estimate.
    The segments denote the edges of the graph and the color is darker if the probability of assigning observations to the edge is greater.
    Right panel: Posterior probabilities of	co-clustering of observations assigned to vertices.
    \label{fig: results mice}}
\end{figure}

\begin{figure}[ht]
    \centering
    \includegraphics[width=0.5\linewidth]{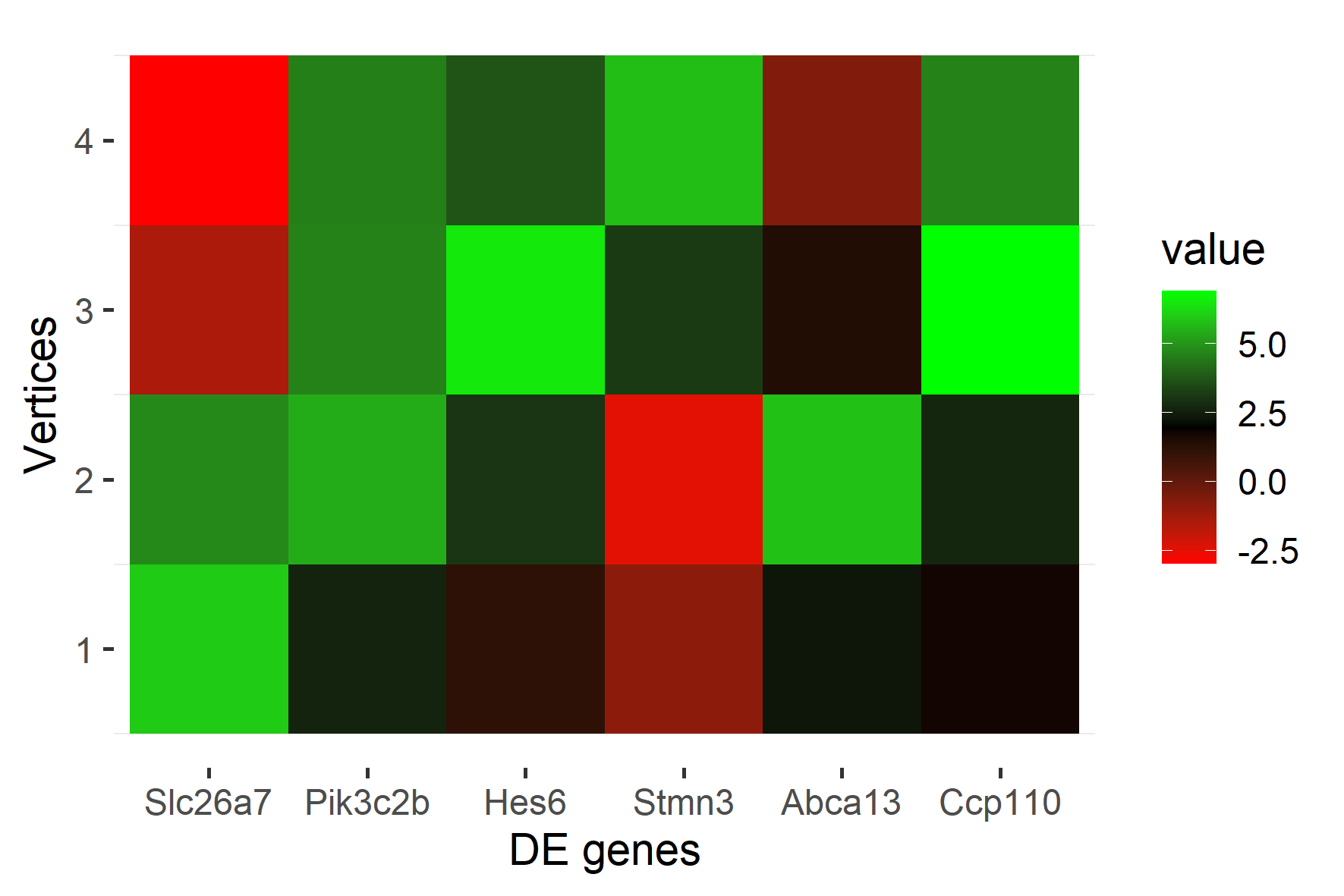}
    \caption{Heatmap of the log mean expressions in top 6 DE genes in the main phases.
    \label{fig: heatmap subset}}
\end{figure}

\begin{figure}[ht]
    \centering
    \includegraphics[width=0.3\linewidth]{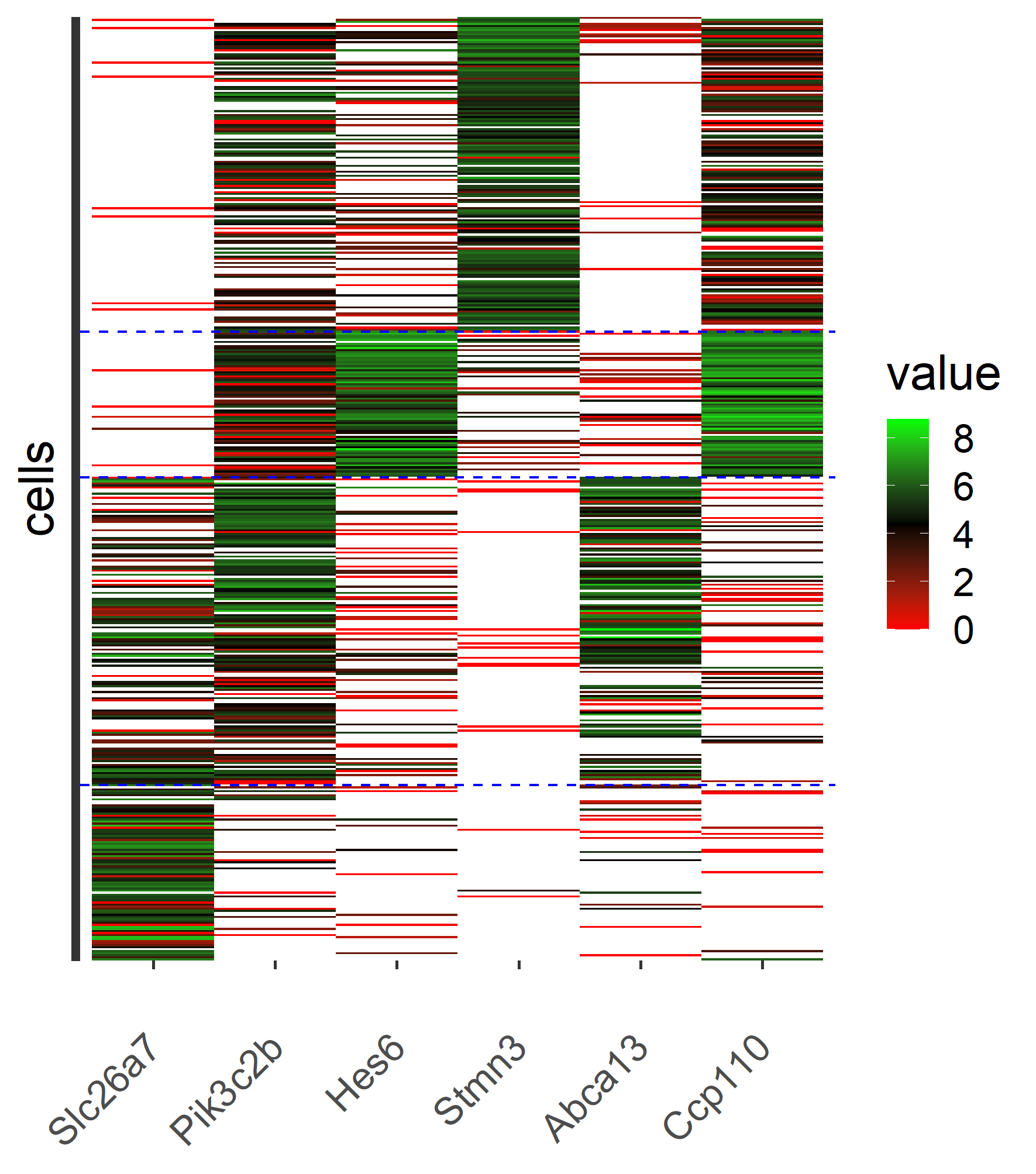}
    \includegraphics[width=0.3\linewidth]{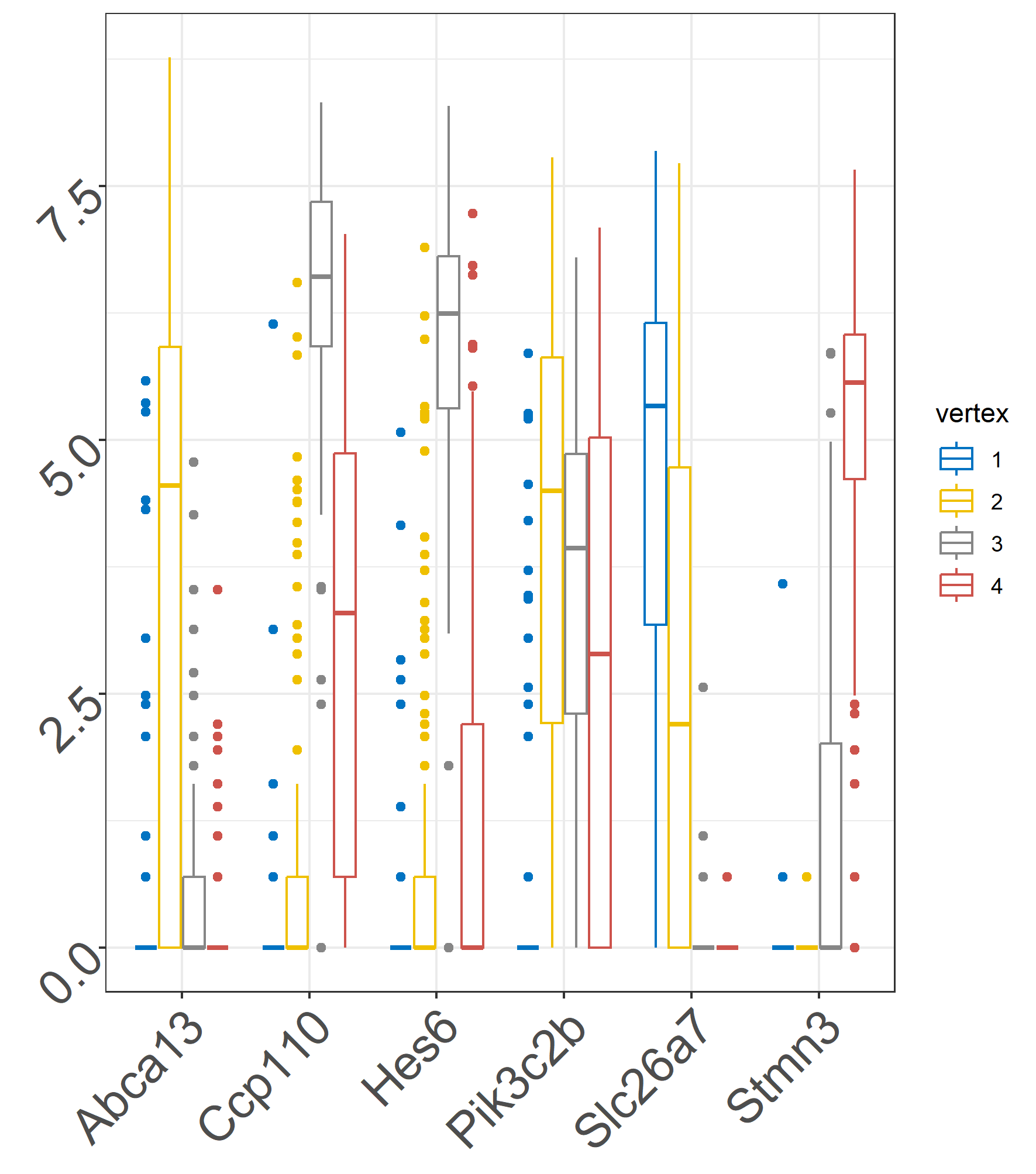}
    \caption{Left panel: Heat-map log genetic expressions in top 6 DE genes in all cells ordered by main phases.
    The cells are sorted by vertex-cluster memberships and the dashed blue lines separate the cells in the different clusters. 
    Right panel: Boxplot genetic expressions (after $\log(\cdot+1)$ transformation) in the top 6 DE genes in all cells in the different main phases (vertex-clusters). 	
    \label{fig: heatmap all cells DE}}
\end{figure}

The posterior estimated GARP places $466$ cells into vertex-clusters (main phases) and $281$ into ordered edge-clusters (transition phases).
Figure \ref{fig: results mice}a summarizes inference.
The heat-map in Figure \ref{fig: results mice}b shows the posterior probabilities of co-clustering of pairs of observations, suggesting low posterior uncertainty around the estimated main phases, making the point estimate under the GARP a meaningful posterior summary.
The conditional uncertainty of the graph-alignment of the vertices given the point estimates of the main phases is low.
Visual inspection of the results suggests that the model is effectively working as expected.
Once we have identified the main phases (vertex-clusters) we find the biomarkers that best characterize such clusters, i.e., the most differently expressed genes (DE genes).
We rely on the function \texttt{findMarkers} of the \texttt{Bioconductor} package \texttt{scran} \citep{lun2016step}.
More precisely, we first perform an exact binomial test to identify DE genes between pairs of groups of cells (vertex-clusters).
From that, we identify the 6 most significant biomarkers for each pairwise comparison.
For each gene then a combined p-value is computed using Simes multiplicity adjustment applied to all p-values obtained by the pairwise comparisons \citep{simes1986improved}. 
Note that these p-values are not directly used for ranking and are only used to find the DE genes. 
Finally, the p-values are consolidated across all genes using the BH method of \cite{benjamini1995controlling} to implement multiple comparisons under a restriction on false discovery rate (FDR) \citep{benjamini2009selective}. 
The adjusted p-values are reported in Table \ref{tab: genes DE}.
The reported FDRs are intended only as a rough measure of significance. 
Note that properly correcting for multiple testing is not generally possible when clusters are based on the same data that is used for the DE testing.
Nonetheless, a small FDR remains desirable. 
Table \ref{tab: genes DE} shows the average within vertex-cluster gene expressions for the selected top $6$ biomarkers and corresponding FDRs.
The log means expression in the different biomarkers and vertices are also shown in Figure \ref{fig: heatmap subset}.
On average the main phases obtained (vertex-clusters) have very different expressions of the selected biomarkers.
Finally, we show the entire distribution of the cells in the different biomarkers and main phases in Figure \ref{fig: heatmap all cells DE}.

\begin{table}[ht]
    \centering
    \begin{tabular}{rrrrrr}
	\hline
	DE Genes & Vertex 1 & Vertex 2 & Vertex 3 & Vertex 4 & FDR \\ 
	\hline
	Slc26a7 & 408.68 & 120.96 & 0.24 & 0.05 & 1.30e-23 \\ 
    Pik3c2b & 14.15 & 231.82 & 105.74 & 98.38 & 1.38e-08 \\ 
    Hes6 & 3.10 & 21.16 & 691.19 & 41.62 & 4.17e-13 \\ 
    Stmn3 & 0.43 & 0.09 & 23.90 & 320.38 & 1.58e-20 \\ 
    Abca13 & 10.15 & 337.45 & 4.25 & 0.54 & 8.29e-08 \\ 
    Ccp110 & 5.75 & 15.19 & 1008.68 & 105.50 & 7.47e-13 \\ 
	\hline
    \end{tabular}
    \caption{Average within vertex-cluster gene expressions and FDRs in the selected top 6 biomarkers\label{tab: genes DE}.}
\end{table}

\subsection{Comparison with Independent Gaussian Mixtures} 
\begin{figure}[ht]
    \centering
    \includegraphics[width=8cm]{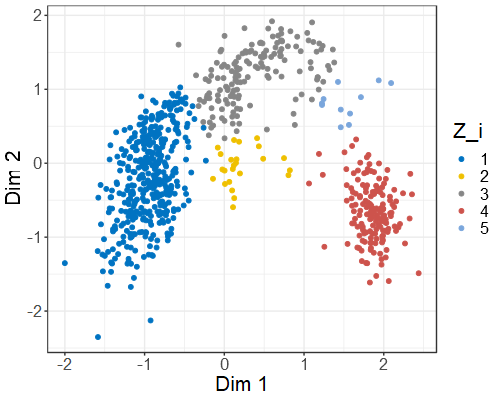}
    \includegraphics[width=8cm]{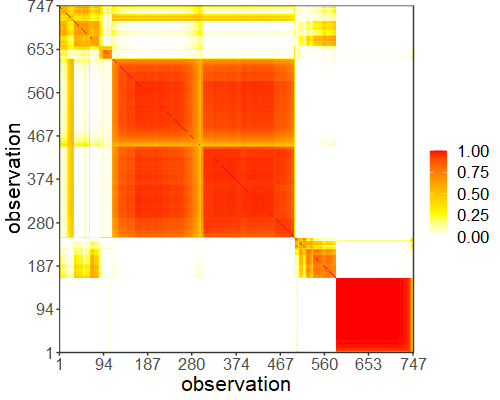}
    \\
    (a) \hspace{7.7cm} (b) \hspace{0.5cm}
    \caption{ Results with independent mixtures. 
    Left Panel: Scatter-plot of the scRNA data. 
    Cells are colored according to the different phases in the point estimate. 
    Right panel: Co-clustering posterior probabilities.	
    \label{fig: results mice orange}}
\end{figure}

For comparison, we estimate an independent Gaussian mixture model without edges and cluster alignment (implemented as the GARP model with $p_{v}=1$).
The posterior distribution of the number of clusters (see Table \ref{tab: prob K_v comparison}) shows more uncertainty since the model fails to find well-separated clusters, due to the noise that is introduced by the presence of the cells transitioning between the main phases.
In other words, including cells in transition in the clustering has reduced the statistical power in detecting homogeneous subpopulations.  
This is illustrated in Figure \ref{fig: results mice orange}.
Recall that we are using variation of information (VI) loss to summarize the posterior random partition.
As a consequence of the increased uncertainty, the point estimate of the clustering of the main phases becomes sensitive to the choice of the loss function.
For instance, both the point estimate and the maximum a posteriori estimate of the number of main phases is $4$ under GARP, while the earlier is $5$ and the latter is $6$ under the independent Gaussian mixture model. 
In the figures, we show the estimated cluster arrangement that minimizes the VI loss for coherency in the comparison.
\begin{table}[ht]
    \centering
    \begin{tabular}{rrrrrrr}
	\multicolumn{7}{c}{Panel A: GARP}\\
	\hline
	$\kv$ & 4 & 5 & 6 & 7 & 8 \\ 
	\hline
	$\hat{P}(K_{v}=\kv)$ & 0.7801 & 0.1951 & 0.0240 & 0.0004 & 0.0004  \\ 
	\hline
    \end{tabular}
    \begin{tabular}{rrrrrrrrrr}
	\multicolumn{10}{c}{Panel B: Independent Mixture Model}\\
	\hline
	$k$            & 5     & 6     & 7     & 8    & 9     & 10   & 11    & 12    & 13 \\ 
	\hline
	$\hat{P}(K=k)$ & 0.0692 & 0.4362 & 0.3115 & 0.114 & 0.0516 & 0.0128 & 0.002 & 0.0012 & 0.0016\\
	\hline
    \end{tabular}
    \caption{Panel A: Estimated posterior of the number of main phases under GARP.\\
    Panel B: Estimated posterior of the number of main phases under the independent Gaussian mixture model.
    \label{tab: prob K_v comparison}}
\end{table}

\FloatBarrier
\section{Discussion} \label{sec: discussion}
We proposed a graph-aligned random partition model to infer homogeneous subgroups of observations aligned on a graph, explicitly allowing for units transitioning between the clusters.
The motivating applications are single-cell RNA experiments where scientists are interested in understanding fundamental biological processes such as cell differentiation and tumor evolution.
Interesting future applications include inference for cell type transitions in a tumor microenvironment. 
Other extensions could include data integration with other modalities, such as histology data.

Methodological extensions include jointly clustering similar cells {\em and} genes, via separately exchangeable nested random partition models \citep{lee2013nonparametric,lin2024separate}.
Another interesting extension is to combine the results of partially exchangeable random partition models that arise from the compositions of Gibbs-type and species sampling priors \citep{teh2006hierarchical,camerlenghi2019distribution,argiento2020hierarchical,bassetti2020hierarchical,lijoi2023flexible} to the GARP model with dependent locations.
In the context of the scRNA-seq experiment, this would allow inference on multiple single-cell RNA-seq data matrices.
In such a way one could borrow information across different measurements while accounting for relevant heterogeneity.
Finally, including unit-specific spatial information, the model can be used for spatial clustering with transitions between the clusters.

\section*{Acknowledgment}
Both authors have been partially supported by NSF/DMS 1952679.
\if1\blind
Most of the paper was completed while G.\ R.\ was a Postdoc at UT Austin.
G.\ R.\ is also affiliated to ``de Castro'' Statistics Initiative, Collegio Carlo Alberto, Torino and acknowledges support of MUR - Prin 2022 - Grant no.\ 2022CLTYP4, funded by the European Union – Next Generation EU.
\fi

\FloatBarrier
\bibliographystyle{natbib}
\bibliography{bib/BNP_cit.bib, bib/Rebaudo_pub.bib}

\clearpage\pagebreak\newpage

\setcounter{equation}{0}
\setcounter{page}{1}
\setcounter{table}{0}
\setcounter{figure}{0}
\setcounter{section}{0}
\numberwithin{table}{section}
\renewcommand{\theequation}{S.\arabic{equation}}
\renewcommand{\thesection}{S.\arabic{section}}
\renewcommand{\thesubsection}{S.\arabic{section}.\arabic{subsection}}
\renewcommand{\theThm}{S.\arabic{Thm}}
\renewcommand{\theCor}{S.\arabic{Cor}}
\renewcommand{\theProp}{S.\arabic{Prop}}
\renewcommand{\theLem}{S.\arabic{Lem}}
\renewcommand{\thepage}{S.\arabic{page}}
\renewcommand{\thetable}{S.\arabic{table}}
\renewcommand{\thefigure}{S.\arabic{figure}}

\if1\blind
{
\bigskip
\begin{center}
{\Large Supplementary materials of \\
			\bf	\mbox{Graph-Aligned Random Partition Model} (GARP)
}
\end{center}
\vskip 2mm
\begin{center}
\spacingset{1}
	\small
	Giovanni Rebaudo$^{a}$ (giovanni.rebaudo@unito.it) \\
	Peter M\"uller$^{b}$ (pmueller@math.utexas.edu) \\
	\vskip 3mm
    $^{a}$
    University of Torino, IT
  	\vskip 4pt 
	$^{b}$
 University of Texas at Austin, USA
\spacingset{1.2}
\end{center}
} \fi

\if0\blind
{
  \bigskip
  \bigskip
  \bigskip
\begin{center}
{\Large Supplementary materials of \\
\bf	\mbox{Graph-Aligned Random Partition Model} (GARP)
}
\end{center}
  \medskip
} \fi

\section{Edge Multivariate Gaussian Mixtures} \label{sec: mv edge Gauss supp}
\begin{figure}[ht] 
    \centering
    \includegraphics[width=0.5\linewidth]{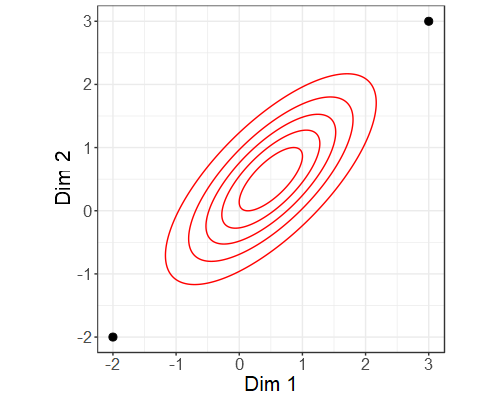}
    \caption{Elliptical contour plot for an edge-cluster (with hyperparameters, as described in Section \ref{sec: hyper supp}), connecting two vertex-clusters $k$ and $\kvp$ (with locations $\mub_{1}=(-2,-2)$ and $\mub_{2}=(3,3)$).
    The vertices are shown as black bullets located on the contour (not shown) line of the bivariate Gaussian such that 99\% of the probability is inside such an ellipse.}
    \label{fig: Gauss edge contour supp}
\end{figure}

Figure \ref{fig: Gauss edge contour supp} shows the contour plot of an edge cluster in $\Re^{2}$.

Without loss of generality consider an edge connecting the two vertex-clusters, $k=1$ and $\kvp=2$, with cluster-specific parameters $\mub_{1}^\ast$ and $\mub_{2}^\ast$.
The edge-cluster is centered around the half-point $\mub_{1,2}^\ast = \frac{\mub_{1}^\ast+\mub_{2}^\ast}{2}$. 
The following construction defines $\Sigbs_{1,2}$ such that the edge is aligned along the connecting line $L_{1,2}$, as described in Section \ref{sec: vert edge} of the main manuscript.
Let $\eb = \frac{\mub_{1}^\ast-\mub_{2}^\ast}{||\mub_{1}^\ast-\mub_{2}^\ast||}$, where $||\mub_{1}^\ast-\mub_{2}^\ast||$ denotes the Euclidean distance between $\mub_{1}^\ast$ and $\mub_{2}^\ast$. 
Let $\Pb=\eb \eb^\intercal$ be the perpendicular projection matrix such that for any $\yb_{i} \in \mathbb{R}^p$, $\yb_{i}^{(p)} = \Pb \yb_{i}$ is the perpendicular projection of $\yb_{i}$ onto the connecting line between $\mub_{1}^\ast$ and $\mub_{2}^\ast$.
Let $\Pb=\Qb\Db\Qb^\intercal$ denote a singular value decomposition (SVD) with $\Db=\text{diag}(1,0,\ldots,0)$.
Thus $\tilde{\Rb}=\Qb^\intercal$ is the rotation matrix such that $\tilde{\yb}_{i}=\tilde{\Rb}\yb_{i}$ is the rotation of $\yb_{i}$ in the new axes where the first axis is the line connecting $\mub_{1}^\ast$ and $\mub_{2}^\ast$ and the others are the orthogonal directions.
Now, we define $\tilde{\Sb} = \text{diag}(||\mub_{1}^\ast-\mub_{2}^\ast|| r_{0}, r_{1}, \ldots, r_{1})$ and $\Sigbs_{1,2} = \Rbt \Sbt \Rbt^{\intercal}$. 

Under this construction, the term in the mixture of normal sampling model \eqref{eq: likelihood} corresponding to the edge $(\kv,\kvp)$ is such that the Gaussian component projected onto the connecting line $L_{1,2}$ has a standard deviation $r_{0} \, ||\mub_{1}^\ast-\mub_{2}^\ast||$, implying lower likelihood for edges between distant vertices' locations. 
The standard deviations of the independent Gaussian distributions on the projection onto $L_{1,2}^\perp$ is $r_{1}$. 

\section{Composition of Discrete Random Probabilities} \label{sec: discr supp}
Let $\thb_{i} = \thbs_{Z_{i}}$ denote the normal moments in the sampling model \eqref{eq: likelihood}. 
As a third characterization of the proposed GARP, we define $\gnt$ as a graph-aligned random partition (with unique atoms) implied by the ties under conditional i.i.d.\ sampling of $\thb_{i}$, with separate models for vertex and edge-clusters. 
For vertex-clusters
\begin{align} \label{eq: prob vert}
    \begin{split}
	&\thb_{i} \mid V_{i} =1, \Vb, P_{v} \iidsim P_{v},\\
        &P_{v} = \sum_{m=1}^{M_{v}} \pi_{m} \delta_{\thbt_{m}} \sim \text{Gibbs-Type Process},
    \end{split}
\end{align}
where $M_{v}$ is the number of atoms of the discrete random probability $P_{v}$ that is a Gibbs-type process and can be finite, as in the finite symmetric DM case, infinite as in the DP, and PYP case, or be a random variable on $\mathbb{N}$ as in the MFM case.
Thus $(\pi_{m})_{m=1}^{M_{v}}$ are the random weights (that are sampled independently from the atoms) from the distribution on the simplex associated with the Gibbs-type process.
The unique atoms $\thbt_{m}$ of $P_{v}$ are i.i.d.\ samples from the NIW distribution in \eqref{eq: prior NIW}.
Note that the unique sampled vertex parameters $\thbs_{v} = \{\thbs_{1}, \ldots, \thbs_{K_{v}}\}$ are a subset of $\{\tilde{\thb_{1}},\ldots, \tilde{\thb}_{M_{v}}\}$. 

The edge-clusters are implied by 
\begin{align} \label{eq: prob edge}
    \begin{split}
        & \thb_{i} \mid V_{i} =0, \Vb\ii, K_{v}, \thbs_{v}, \Tr \iidsim P_{e},\\
        & P_{e}= \sum_{1 \le \kv<\kvp\le K_{v}} \pi_{\kv,\kvp} \delta_{(\thbs_{\kv,\kvp})}.
    \end{split}
\end{align}
Recall that $M_{e}=K_{v}(K_{v}-1)/2$.
The random weights follow a symmetric $M_{e}$-dimensional Dirichlet with hyper-parameter $\beta/M_{e}$, 
\begin{equation} \label{eq: weight edge}
    (\pi_{\kv,\kvp})_{1\le \kv < \kvp \le K_{v}} \sim \text{Dir}(\beta/M_{e}, \ldots, \beta/M_{e}).
\end{equation}
Finally, recall that \eqref{eq: prior V_i} might generate $N_{e}>0$, even when \eqref{eq: Gibbs urn} implies $M_{e}=0$. 
For this case, we define for completeness $\tildegn
\{\thb_{i}=(\mathbf{0}, I_{p}) \mid V_{i}=0, \Zb\ii, \Tr^{c}\} \equiv
1$, where $\mathbf{0}$ is a $p$-dimensional vector of $0$'s and $I_p$
is a $p\times p$-dimensional identity matrix (without implications for
$\gn$, due to the truncation to $\Tr$ in \eqref{eq: trunc urn}).

From the characterizations of Gibbs-type and DM processes, it is straightforward to show that the aforementioned discrete conditional random probability models for the parameters characterize the GARP as stated in the following proposition.

\begin{Prop} \label{prop: GARP vs rand prob}
    The random partition structure of the {\rm{GARP}} model \eqref{eq: pr V Z} and the vertex- and edge-parameters distributions can be characterized as the configuration of ties implied by the truncation sampling model in \eqref{eq: trunc urn}, \eqref{eq: prior V_i}, (S.1), \eqref{eq: prob edge}, and \eqref{eq: weight edge}.
\end{Prop}

\section{Hyperparameters Settings\label{sec: hyper supp}}
In both the application and the simulation we set $\gamma=0.5$ for the Gnedin process controlling the vertex-clusters and $\beta=0.5$ for the symmetric DM with hyperparameter $0.5/M_{v}$ to favor the sparsity of the graph.
Moreover, for the choice of the hyperparameters of the NIW we set $\mub_{0}= \bar{\yb}$, $\kappa_{0} = 0.001$, $\nu_{0} = 100$, $\mathbf{\Lambda}_{0} = \xi^2 \, \mathbf{I}$ and $\Sigb_{0} = \mathbf{\Lambda}_{0}^{-1}$.
For scenarios in which the clusters are well separated, we recommend a large value of $\xi^2$ (that we set equal to $150$), while we recommend a smaller value of $\xi^2$ (that we set equal to $15$) if the data are not well separated in the Euclidean space. 
Moreover, in both the application and the simulation we set $r_{0}^{2}=4 (\chi_{2,1-\alpha}^{2})^{-1}$ and $r_{1}^{2}= (2 \chi_{2,1-\alpha}^{2})^{-1}$, where $\chi_{2,1-\alpha}^{2}$ is the quantile of order $1-\alpha$ (we set $\alpha=1\%$) of a Chi-squared distribution with 2 degrees of freedom to have the desired eccentricity of the elliptical contour plot of the edge as well as the $99\%$-level of the contour plot not too spread. 
To obtain that, recall that $c$-level counter-plot of multivariate Gaussian density, such as the edge Gaussian in \eqref{eq: likelihood}, are points $\yb \in \Re^d$ such that $(\yb-\mub_{\kv,\kvp})^{\intercal} \Sigb_{\kv,\kvp}^{-1} (\yb-\mub_{\kv,\kvp})$ is constant, that is the contour levels are ellipsoid centered at $\mub_{\kv,\kvp}$.
Finally, note that if $\yb\sim \Norm(\mub_{\kv,\kvp}, \Sigb_{\kv,\kvp})$ then, 
$$
(\yb-\mub_{\kv,\kvp})^{\intercal} \Sigb_{\kv,\kvp}^{-1} (\yb-\mub_{\kv,\kvp}) \sim \chi^{2}_{d} \, ,
$$
where $ \chi^{2}_{d}$ denotes the Chi-square distribution with $d$ degrees of freedom.

\begin{figure}[ht]
    \centering
    \includegraphics[width=0.5\linewidth]{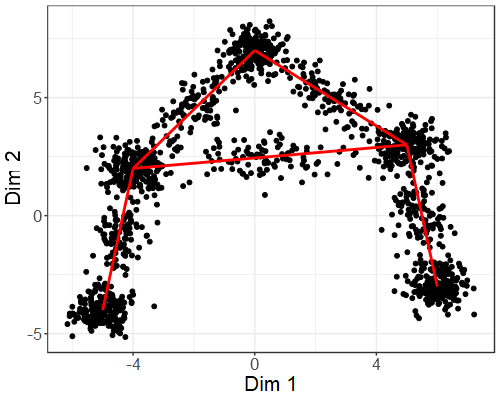}
    \caption{Scatter-plot of the simulated data. 
    The red segments represent the edges that connect the truth vertices.}
    \label{fig: plot simulated data supp}
\end{figure}

Visually the contour plots of such edge pdf are shown in Figure \ref{fig: Gauss edge contour supp} and the data sampled from such configuration looks like the one in Figure \ref{fig: plot simulated data supp}.

\section{Implementing Posterior Inference}
\subsection[Use of the Relaxed Model in Posterior Simulation]{Use of the Relaxed Model $\tildegn$ in Posterior Simulation}
\label{sec: neal}
We discuss in more detail the use of the projectivity property of $\tildegn$ to define a P\'olya urn scheme for a tractable marginal posterior simulation algorithm.
First, recall that the relaxed model $\tildegn$ can be seen as a hierarchical composition of a Kolmogorov consistent EPPFs with product partition forms (Sections \ref{sec: GARP} and \ref{sec: comp Gibbs}), which implies tractable expressions for $\tildegn(V_i, Z_i \mid \Vb^{\ii}, \Zb^{\ii})=\frac{\tildegn(\Vb, \Zb)}{\tildegn(\Vb^{\ii}, \Zb^{\ii})}$ under the relaxed model $\tildegn$.

To derive then the desired full conditional distributions under $\gn$ we note, e.g., that if $I_{i}=0$ for $(\Vb, \Zb)$ (recall the definition of $I_{i}$ in Section \ref{sec: posterior inference}), then
$$
\gn\{V_{i}=v,Z_{i}=z \mid \cdots\} \propto \dfrac{\tildegn(\Vb,\Zb) \prod_{j\in
[N]} \Norm(\yb_{j}\mid \thbs_{Z_{j}})} {\tildegn(\Vb^{\ii},\Zb^{\ii})
\prod_{j\in [N]^{\ii}} \Norm(\yb_{j}\mid \thbs_{Z_{j}})},
$$
for any $v\in \{0,1\}$ and $z \in \Zb^{\ii}$.
Moreover, when $I_{i}=0$ for $(\Vb,\Zb)$, the marginal probability in the denominator is equal to the one in a Kolmogorov consistent model (i.e.,  if $I_{i}=0$, $\tildegn(\Vb^{\ii},\Zb^{\ii})=\widetilde{G^{(N-1)}}(\Vb^{\ii},\Zb^{\ii})$, up to a normalization constant) and this allows us to generalize then tractable marginal samplers such as in Neal (2000) or Teh et al.\ (2006) relying on the characterization of the GARP via a composition of gCRP in Section \ref{sec: gCRP}. 

\subsection{Point Estimates for the GARP Random Partition}
\label{sec: Dahl supp}
How to choose good summaries (i.e., point estimates) for reporting posterior inference on functionals of interest can be a fundamental and nontrivial question in Bayesian analysis.  
It is especially challenging if the object of interest is a partition or a graph.
To define a posterior point estimate and perform uncertainty quantification we build on the existing literature of posterior point estimates of random partition based on a decision-theoretic approach (Wade and Ghahramani, 2018; Dahl et al., 2022b) generalizing the results for the more challenging case of GARP.
We propose a point estimate for the GARP as follows.

\medskip
\texttt{(1)} Assign observations to vertices versus edges using the posterior mode,
\begin{equation*}
    \hat{V}_{i}=1 \text{ if } \Vbar_{i} \equiv \sum_{t} \frac{V_{i}^{(t)}}{T} > 0.5,
\end{equation*}
where $T$ is the Monte Carlo sample size, and $V_{i}^{(t)}$ is the imputed value in iteration $t$ of the MCMC simulations.
The uncertainty around the point estimate is quantified using $(1-\hat{V}_{i})\Vbar_{i} + \hat{V}_{i} (1-\Vbar_{i})$. 

\medskip
\texttt{(2)} Given $\hat{\Vb}$ we find a point estimate $\Zhv$ for the partition of vertex units by minimizing the variation of information loss (VI) (Meil\u{a}, 2007) as suggested by Wade and Ghahramani (2018) and implemented in the \texttt{R} package \texttt{salso} (Dahl et al., 2022a). 
Alternative loss functions can be used as needed for different applications (See e.g., Binder, 1978; Franzolini and Rebaudo, 2023). 
For uncertainty quantification, we report the heat-map with the posterior probabilities of co-clustering. 

\medskip
\texttt{(3)} Given $\hat{\Vb}$ and $\Zhv$ we find a point estimate $\Zhe$ and conditional uncertainty quantification for $\Ze$ using the posterior probability of observations being assigned to the different edges.
We evaluate conditional posterior probabilities of assigning the remaining observations to the possible edges, 
\begin{equation}
    G(\Ze \mid \cdots) \propto \prod_{k<\kvp} \Gamma(n_{\kv,\kvp} + \beta/M_{e}) \prod_{C_{\kv,\kvp}} \Norm(\yb_{i} \mid \mub_{\kv,\kvp}^\ast, \Sigb_{\kv,\kvp}^\ast).
    \label{eq: GZe}
\end{equation}
Here the first product goes over all $(k,\kvp)$ with $1 \le \kv < \kvp \le K_{v}$ and the second over the $\yb_{i}$ such that $z_{i}=(k,\kvp)$, i.e., the set $C_{\kv,\kvp}$.
Probabilities \eqref{eq: GZe} are evaluated by Rao-Blackwellization (Robert and Roberts, 2021), using the full conditionals
\begin{equation*}
    G\{Z_{i}=(\kv,\kvp)\mid V_{i}=0, \cdots\} \propto (n\ii_{\kv,\kvp} +\beta/M_{e}) \text{Norm}(\yb_{i} \mid \mub_{\kv,\kvp}^\ast, \Sigb_{\kv,\kvp}^\ast).
\end{equation*}
We visualize $p(\Ze \mid \Zhv, \, \mub^\ast, \, \Sigb^\ast)$ by adding edges between vertices with color intensity proportional to the sum over observations assigned to edges (i.e., $\hat{V}_i=0$) of the probability that such observations will be assigned to the different edges $(\kv, \kvp)$'s.

\section{Simulation Studies} \label{sec: simulation supp}
We carried out a simulation study under a well-specified and a miss-specified data generating truth to assess inference under finite sample size scenarios.
We set up simulation truths close to the mouse data.
The data are simulated from a 5 vertex mixture with $n_{k}=200$ observations in each vertex and an additional $n_{\kv,\kvp}=100$ observations around 5 assumed edges.

\subsection{Well Specified Scenario} \label{sec: Well Sim}
\begin{figure}[ht]
    \centering
    \includegraphics[width=8cm]{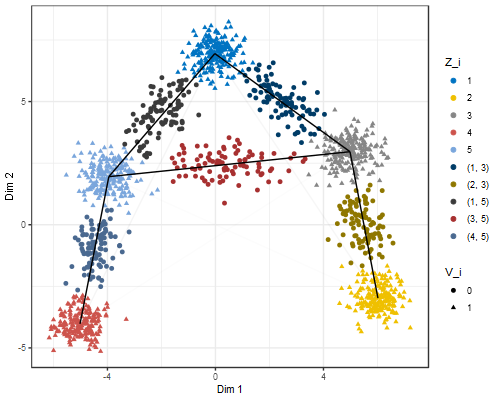}
    \includegraphics[width=8cm]{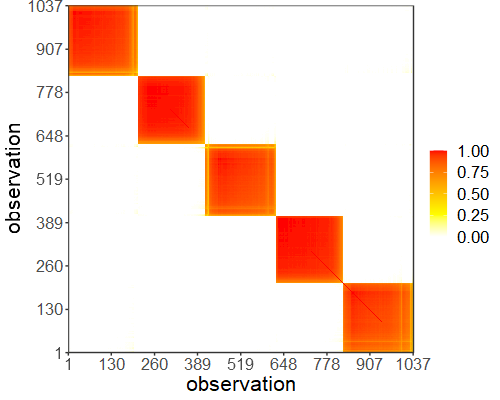}
    \\
    (a) \hspace{7.5cm} (b) \hspace{0.4cm}
    \caption{Well-specified simulation scenario.
      Left Panel: Scatter plot of the simulated data.
      Observations are colored according to the estimated cluster membership. 
      Line segments show edges of the estimated graph, with the clusters at the end of line segments being vertex clusters, and clusters along the line segments being edge clusters.
      The grey level of the line segments shows the estimated probability of assigning observations to the respective edge (barely varying in this case).
      Right panel: Posterior co-clustering probabilities for all observations assigned to vertices.
    \label{fig: results sim}}
\end{figure}

In the first simulation scenario, we assume a simulation truth with $K_v=5$ vertex clusters with cluster-specific Gaussians with mean vectors $(-5,-4)$, $(-4,2)$, $(0,7)$, $(5,3)$ and $(6,-3)$, and a common covariance matrix $\text{diag}(0.25,0.25)$.
Observations assigned to edge components are sampled from a Gaussian mixture with cluster-specific kernels as in \eqref{eq: prior edge atom}.
The simulated $N=1500$ observations are shown in Figure \ref{fig: results sim}a.

Figure \ref{fig: results sim} shows that the GARP was able to recover the simulated truth in the point estimate. 
Moreover, the uncertainty around the point estimate is low.

\subsection{Misspecified Scenario}\label{sec: Miss Sim}
\begin{figure}[ht]
    \centering
    \includegraphics[width=8cm]{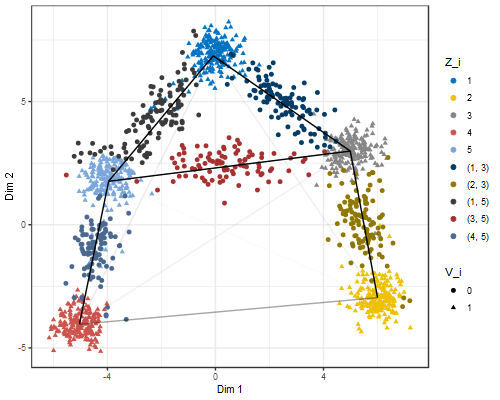}
    \includegraphics[width=8cm]{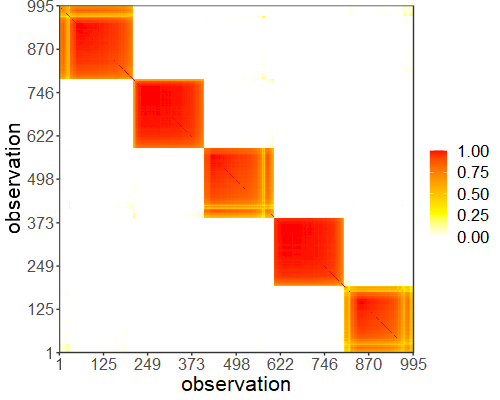}
    \\
    (a) \hspace{7.5cm} (b) \hspace{0.4cm}\\
    \caption{Mis-specified simulation scenario.
      Left Panel: Scatter plot of the simulated data.
      Observations are colored according to estimated cluster membership. 
      The line segments denote estimated edges, with clusters at the end of the line segments being vertex clusters, and clusters along the line segments being edge clusters.
      The gray shade of the line segments indicates the  probability of
      assigning observations to the respective edge. 
      Right panel: Posterior co-clustering probabilities for observations assigned to vertices.
    \label{fig: results sim rev}}
\end{figure}

Here we consider a misspecified data-generating truth, using the same true mean vectors for five vertex clusters with cluster-specific Gaussian kernels as in the previous scenario, but inflated vertex-specific covariance matrices $\text{diag}(0.5,0.5)$.
For the edge components, we introduce two sources of misspecification.
First, we center the edge components not at the midpoint of the two adjacent vertices but introduce a bias term.  
Instead, the edge-specific kernels are centered at $\frac{\mubs_{\kv}+\mubs_{\kvp}}{2}$ plus a shift of $+0.25$ in the direction of the line connecting the adjacent vertices, as well as in the perpendicular direction.
Second, the observations for the edge components are generated from a uniform distribution on a rectangle centered at the described $\mubs_{\kv,\kvp}$ and with the length of the side in the direction of the connecting line equal to half the length of the Euclidean distance between the adjacent vertices and the length of the other side equals $2$.
Under this simulation truth, the scatter plot of the simulated data still allows a meaningful definition of vertex and edge clusters, but the additional misspecification and variability with respect to the well-specified scenario make the inference with our model more challenging.
The simulated $N=1500$ observations are shown in Figure \ref{fig: results sim rev}a.

Figure \ref{fig: results sim rev} shows that the GARP was able to recover well the simulated truth in the point estimate in this misspecified scenario. 
The uncertainty around the point estimate is low.

\subsection{Non-Connected Graph Scenario} \label{sec: Orange Sim}
Here we investigate how the model works in a scenario with no meaningful notion of the connected graph in the data. 
More precisely, we simulate from a mixture of five vertex clusters, exactly as in Section \ref{sec: Well Sim}, but without any edge components.

\begin{figure}[ht]
    \centering
    \includegraphics[width=8cm]{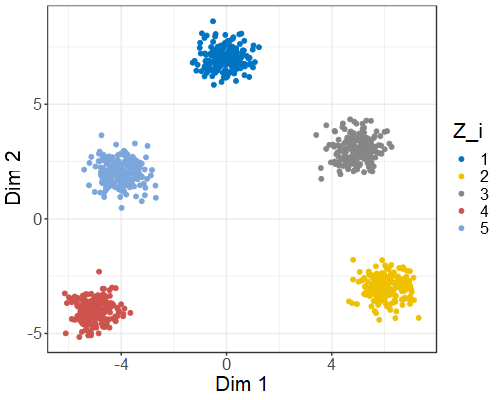}
    \includegraphics[width=8cm]{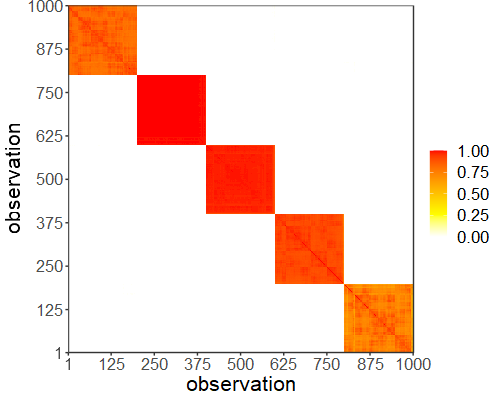}
    \\
    (a) \hspace{7.5cm} (b) \hspace{0.4cm}
    \caption{Non-connected graph simulated data. 
      Left Panel: Scatter plot of the simulated data.
      Observations are colored according to estimated cluster membership.
      Correctly recovering the simulation truth there are no estimated
      edge clusters.
      Right panel: Posterior co-clustering probabilities.
    \label{fig: results sim rev2}}
\end{figure}

The simulated $N=1000$ observations and inference under the GARP are shown in Figure \ref{fig: results sim rev2}a. 

Figure \ref{fig: results sim rev2} shows that the GARP was able to recover well the simulated truth in the point estimate also under this not-connected graph simulation truth. 
\FloatBarrier

\section{Proof of the Main Results}\label{sec: proof supp}

For easy reference, we provide in Table \ref{table: dist supp} a brief statement of the various probability models used in the discussion and the results, and in the following list a brief summary of the main results.
Here, Ex.\ \ref{ex: sym dir} -- \ref{ex: PYP} refer to the four examples for the EPPF from Section \ref{sec: comp Gibbs}.
\medskip

\begin{table}[ht]
    \medskip
    \begin{tabular}{ll}\
        $\gn$ & GARP model \eqref{eq: pr V Z}\\[8pt]
        $\gnt$ & relaxed model \eqref{eq: trunc urn}, $\gnt$ is proportional to $\gn$, but w/o $\Ind(\Tr)$\\[8pt]
        $\gtnn$ & law under $\gnt$ on $(T_{i},\, i=1,\ldots,N)$, $T_{i} = 
        \begin{cases} 
            V_{i} & \mbox{if } V_{i}=0\\
            (V_{i},Z_{i}) & \mbox{if } V_{i}=1
        \end{cases}$\\[8pt]
        $\gt$ & Kolmogorov-consistent extension of $\gtnn$ to $N \in \N$\\[8pt]
        $\ginf$ & inf exch.\ law that eventually matches the predictives under Ex \ref{ex: sym dir} or \ref{ex: GP} (or any MFM)\\[8pt]
        $\ginf_{N}$ & marginal law under $\ginf$
    \end{tabular}
    \caption{Probability models used in the discussion and main results}
    \label{table: dist supp}
  \end{table}
  
For notational simplicity, we refer with $\gt$ also to the marginal laws of the stochastic process $(T_{i})_{i\in \mathbb{N}}$ as well as the law of $M_{v}$ and $(\pi_{m})_{m=1}^{M_{v}}$ in (S.1)  since they do not depend on the dimension $N$.
Finally, we refer with $\gn(\cdot)$ to the probability density and mass functions of random variables under the GARP model \eqref{eq: likelihood}--\eqref{eq: pr V Z}.  
More generally, given a probability measure $P$ we denote by $P\{E\}$ the probability measure evaluated in a measurable set $E$ and by $P(a)$ the corresponding (when it exists) with respect to Lebesgue (i.e., pdf) or counting measure (i.e., pmf) evaluated in a point $a$.

\begin{description}
    \item[Propositions \ref{prop: GARP vs urn} and \ref{prop: GARP vs rand prob}:] Characterizations of the GARP $\gn$ as truncation of $\gnt$, which in turn is characterized as (i) a gCRP or, (ii) a composition of random discrete prob measures, respectively.

    \item[Proposition \ref{prop: prob trunc}:] Analytical statement of $\gnt\{\Tr\}$ for a general Gibbs-type prior.
    
    \item[Theorem \ref{th: prob const}:] Let $g^{\infty}=\lim_{N \rightarrow \infty} \tildegn{\{K_{v}=1\}}$ and $g_{v}^{\infty} = \lim_{n_{v} \rightarrow \infty} \gnt\{K_{v} = 1 \mid N_{v}=n_{v}\}$.
    Then
    $$
    g^{\infty}=g^{\infty}_{v}=
    \begin{cases}
        0 & \mbox{Ex \ref{ex: sym dir} , \ref{ex: DP}, \ref{ex: PYP}}\\
        \gamma \in (0,1) & \mbox{Ex \ref{ex: GP}}
    \end{cases}
    $$
    
    \item[Theorem \ref{th: const eventually}:] $\gt\{ \Tr \text{ eventually} \} = 
    \begin{cases}
        1 & \mbox{Ex \ref{ex: sym dir} , \ref{ex: DP}, \ref{ex: PYP}}\\
        1-\gt\{M_{v}=1\} & \mbox{Ex \ref{ex: GP}}
    \end{cases}$.
    
    \item[Proposition \ref{prop: fEPPF}:] $\text{fEPPF}^{(N)}_{K}$ under the GARP model in \eqref{eq: pr V Z}.

    \item[Proposition \ref{prop: fin exch no inf}:] The data $\yb $, the graph-aligned random partition induced by $(V_{i}, Z_{i})$ and the random partition $\Psi_{N}$ are finitely exchangeable, but not a projection of an infinitely exchangeable process under our proposal \eqref{eq: likelihood}--\eqref{eq: pr V Z}.
    
    \item[Theorem \ref{th: limit de Fin Dir} and Corollary \ref{cor: limit MFM}:] Under Ex \ref{ex: sym dir}, the predictive probabilities for $V_{i}, Z_{i}$ under the GARP are eventually equal to the same under a Kolmogorov-consistent sequence $\big(\ginf_{N}\big)$; statement of a P\'olya urn and directing measure for $\big(\ginf_{N}\big)$.\\[4pt]
    The same remains true for any MFM.
\end{description}

\subsection{Proof of Proposition \ref{prop: GARP vs urn}}
\begin{proof}
We assume the GARP definition via the \emph{relaxed model} in \eqref{eq: trunc urn}, \eqref{eq: prior V_i}, \eqref{eq: Gibbs urn}, and \eqref{eq: DM urn} and show that is equivalent to the definition in \eqref{eq: pr V Z}.  
	
First, we note that in \eqref{eq: pr V Z} the constraint $\Ind(\Tr)$ can be rewritten as $\Ind(\{N_{v}=N\} \cup \{K_{v} > 1\})$.
Note also that under $N_{e}=0$ the second line in \eqref{eq: pr V Z} does not arise.
For notational simplicity, we naturally extend the definition of $K_{v}$ and $M_{e}$ by defining $K_{v}=M_{e}=0$ if $N_{v}=0$ and defining   $\DM_{\cdot}^{(0)}(\cdot) = \DM_{0}^{(\cdot)}(\cdot) \equiv 1$. 
	
Note also that \eqref{eq: Gibbs urn} is equivalent to sample $\Zbv= (Z_{i}: i \in [N], \, V_{i}=1)$ from 
\begin{equation} \label{eq: Zv supp}
    \tildegn(\Zv \mid \Vb)= \EPPF_{K_{v}}^{(N_{v})}(n_{1},\ldots,n_{K_{v}}\mid \alpha, \sigma)/K_{v}!
\end{equation}
The clustering indicators $Z_{i}$ are a 1-to-1 mapping of the induced exchangeable random partition up to possible relabelings. 
By (conditional) exchangeability of the partition, any possible relabeling of $\Zv$ has the same probability that is equal to the EPPF divided by the number of relabelings, i.e., $K_{v}!$. 
	
Similarly, sampling from \eqref{eq: DM urn} is equivalent to	sample $\Ze=(Z_{i}: i \in [N], \, V_{i}=0)$ from
\begin{align*}
    \tildegn(\Ze \mid \Zv) = \DM_{M_{e}}^{(N_{e})}((n_{\kv,\kvp})_{\kv < k_{v}^{\prime}} \mid \beta/ M_{e} ),
\end{align*}
where $\DM_{M_{e}}^{(N_{e})}((n_{\kv,\kvp})_{\kv < k_{v}^{\prime}})$ denotes the marginal likelihood of the DM distribution for the categorical random variables, which is a function of the sufficient statistics $(n_{\kv,\kvp})_{\kv < k_{v}^{\prime}}$, i.e., the ordered cardinalities of the different edges.
In contrast to the EPPF and fEPPF, here some $n_{\kv,\kvp}$ can be $0$, implying that there is no edge connecting the vertices $\kv$ and $\kvp$.
	
Finally, we obtain \eqref{eq: pr V Z} via   the multiplication rule of probability, i.e., 
$$
\tildegn(\Vb,\Zb) = \tildegn(\Vb) \, \tildegn(\Zv \mid \Vb) \cdot \tildegn(\Ze \mid \Vb, \Zv)
$$
where $\tildegn(\Vb)=p_{v}^{N_{v}}(1-p_{v})^{N_{e}}$ by \eqref{eq: prior V_i}.
\end{proof}

\subsection{Proof of Proposition \ref{prop: prob trunc}}
\begin{proof}
First, recall $\Tr = \{N_{v}=N\} \cup \{ K_{v} > 1\}$.
That is, $\Tr$ occurs if and only if there are at least two vertex-clusters (i.e., $K_{v}>1$) unless no observations are allocated to edge-clusters (i.e., $N_{v}=N$).
Thus, by additivity of probability,
\begin{align*}
    \tildegn\{\Tr\} &= \tildegn\{\{N_{v}=N\} \cup \{K_{v} > 1\}\} \\
                    &= \tildegn\{N_{v}=N\} + \tildegn\{\{N_{v} \ne N\} \cap \, \{K_{v} > 1\}\},
\end{align*}
where $\tildegn\{N_{v}=N\} = p_{v}^{N}$.
In words, we decompose $\Tr$ into the union of the (disjoint) events ``all clusters are vertices'' and ``not all observations are in vertices and there are at least 2 vertex-clusters''.
The second term is further expanded by conditioning on $N_{v}$ as:
\begin{align*}
    &\tildegn\{\{N_{v} \ne N\} \cap \{K_{v} > 1\}\} =\\
    & \tildegn\{\{N_{v} \notin \{0,1,N\}\} \cap \{K_{v} > 1\} \} =\\
    & \sum_{n_{v}=2}^{N-1} \tildegn\{N_{v}=n_{v}\} \tildegn\{K_{v} \ne 1 \mid N_{v}=n_{v}\}=\\
    & \sum_{n_{v}=2}^{N-1} \binom{N}{n_{v}} p_{v}^{n_{v}} (1-p_{v})^{n_{v}-1} \big[1-\EPPF_{1}^{(n_{v})}(n_{v}) \big]=\\
    & \sum_{n_{v}=2}^{N-1} \binom{N}{n_{v}} p_{v}^{n_{v}} (1-p_{v})^{n_{v}-1} \big[1-(1-\sigma)_{n_{v}-1} W_{n_{v},1} \big],
\end{align*}
where the last equality follows from the definition of the Gibbs-type priors.
\end{proof}

\subsection{Proof of Theorem \ref{th: prob const}}
\begin{proof}
First, note that the finite sample behavior of $$g_{n_{v}}=\tildegn\{K_{v, N}=1\mid N_{v, N}=n_{v}\} =\gt\{K_{v, N}=1\mid N_{v,N}=n_{v}\} =\EPPF_{1}^{(n_{v})}(n_{v})$$ is derived as a special case of the EPPF in the different examples in Section \ref{sec: comp Gibbs} of the main manuscript.
From it, we can derive the large sample behavior $g_{n_{v}}$ and the limit $g_{v}^{\infty}$ reported in Table \ref{tab: g^inf}.
Let $(x)_{n}=\Gamma(x+n)/\Gamma(x)=x(x+1)\cdots(x+n-1)$.  
To compute the rate of $g_{n_{v}}$ we note that by the Stirling approximation
$$\frac{(x)_{n}}{n!} = \frac{\Gamma(x+n)}{\Gamma(x) n!} \asymp \frac{n^{x-1}}{\Gamma(x)} \quad \text{as } n \rightarrow \infty.$$

Note also that $(N_{v,N})_{N\in \mathbb{N}}$ is a ($\gt$-almost surely) Markovian non-decreasing sequence of random integers such that
$$\frac{N_{v,N}}{N} \rightarrow p \quad \text{as } N \rightarrow \infty$$ 
$\gt$-a.s.\ by the strong law of large numbers.
Therefore, $N_{v,N}$ diverges $\gt$-almost surely and $g_{v}^{\infty} \equiv \lim_{n_{v}\to\infty} \gt\{K_{v}=1\mid N_{v,N}=n_{v}\} = \lim_{N\to\infty} \gt\{K_{v}=1\} = g^{\infty}.$

We note, as a remark, that to have $g_{v}^{\infty}$ well defined we consider a sequence $(N=f(n_{v}))_{n_{v} \in \N}$ such that $f:\mathbb{N} \rightarrow \mathbb{N}$ and $f(n) \ge n$ for any $n \in \mathbb{N}$.
Moreover, the hierarchical definitions of $\mathbf{V}$ and $\Zv$ imply that $K_{v}= K_{v}(\Zb_{v})$ $\gt$-almost surely, where $K_{v}= K_{v}(\Zb_{v})$ indicates a function of the $N$ units $(V_{i},Z_{i})$ that depends on $\Zb$ only indirectly through the $N_{v,N}$ units allocated to vertices, i.e., $\Zb_{v}$. 

Finally, as derived in Section \ref{sec: understanding} of the main manuscript, $g_{v}^{\infty}=g^{\infty}=\lim_{N \rightarrow \infty} \gt\{\Tr^{c}\}$. 
\end{proof}

\subsection{Proof of Theorem \ref{th: const eventually}}

Recall the definition of eventually.
Let $(\Tr)_{N\in \mathbb{N}}$ be a sequence of events in the measurable space $(\Omega, \mathcal{F})$,
$$\{\Tr \text{ eventually} \} = \liminf_{N} \Tr = \cup_{\bar{N}=0}^{\infty} \cap_{N=\bar{N}}^{\infty} \Tr.$$
In words, it is the set of $\omega \in \Omega$ such that there exists an integer $\bar{N}(\omega)$ such that for any integer $N \ge \bar{N}(\omega)$, $\omega \in \Tr$.
\begin{proof}
{\bf Case with a $M_{v}$-dimensional symmetric Dirichlet (where $M_{v}>1$) or with a DP or with a PYP in \eqref{eq: pr V Z}.}
\smallskip
\newline
First, since $K_{v,N}, N_{v,N}$ are functions of $T_{1:N}$ (that is of $(\Vb_{1:N}, \Zb_{v,N})$) only, $\tildegn(K_{v,N},N_{v,N})=\gt(K_{v,N},N_{v,N})$ for any $N \in \mathbb{N}$.

Note that under $\gt$, $(K_{v, N})_{N}$ is an a.s.\ non-decreasing Markovian sequence of positive integers such that for any natural $N > 1$, $\gt \{K_{v, N} > 1\} >0$ and it can be computed from \eqref{eq: prior V_i}-\eqref{eq: Gibbs urn}.

Moreover, by Kingman's representation theorem (see Kingman, 1978 and Theorem 14.7 in Ghosal and van der Vaart, 2017) the random partition can be characterized as arising from the ties obtained by sampling from a unique discrete probability measure $P_{v}=\sum_{m=1}^{M_{v}} \pi_{m} \delta_{\tilde{\thb}}$ (we know that is $M_{v}$-symmetric Dirichlet or a DP or a PYP distributed) and the frequency of the $k$th largest partition block converges almost surely to $k$th largest random weight in $(\pi_{m})_{m=1}^{M_{v}}$ for any $k \in 1,\ldots, M_{v}$.
Therefore, together with the assumption $M_{v} \ge 2$, it implies that
$$ \gt\{ \{K_{v,N} > 1\}\text{ eventually } w.r.t.\ N\}=1.$$
To conclude the proof of \eqref{eq: eventually E_n}, note that 
$$
E_{N} = \{N_{v,N}=N\} \cup \{K_{v,N}>1\} \supset \{ K_{v,N}>1\}.
$$
Thus, we have shown that $\gt\{\Tr \mbox{ eventually}\}=1$.

To prove \eqref{eq: fin Dir tilde pred}, first recall that for any $N \in \mathbb{N}$, $\gn$ and $\gnt$ denote the probability mass function of $(V_{i}, Z_{i})_{i=1}^{N}$ under the GARP and the relaxed model, respectively. 
Next, for any $N, k \in \mathbb{N}$ and any set of possible points $a_{k} = (\mathbf{v}_{1:N+k}, \mathbf{z}_{1:N+k})$, by definition of conditional probability we have
\begin{equation}
    \widetilde{G^{(N+k)}}(a_{k} \mid \mathbf{V}_{1:N}=\mathbf{v}_{1:N},\mathbf{Z}_{v,N}=\mathbf{z}_{v,N})= \frac{\widetilde{G^{(N+k)}}\big(a_{k} \big)}{\widetilde{G^{(N+k)}}\{ \mathbf{V}_{1:N}=\mathbf{v}_{1:N},\mathbf{Z}_{v,N}=\mathbf{z}_{v,N}\}},
\label{eq: cond}
\end{equation}
where, by additivity of probability,
$$
    \widetilde{G^{(N+k)}}\{ \mathbf{V}_{1:N}=\mathbf{v}_{1:N},\mathbf{Z}_{v,N}=\mathbf{z}_{v,N}\} = \sum_{\big\{(v_{i}^{\prime}, z_{i}^{\prime})_{i=1}^{N+k} : \mathbf{v}_{1:N}^{\prime}=\mathbf{v}_{1:N}, \, \mathbf{z}_{v,N}^{\prime} =\mathbf{z}_{v,N} \big\}} \widetilde{G^{(N+k)}}((v_{i}^{\prime}, z_{i}^{\prime})_{i=1}^{N+k} ).
$$
Moreover, for any $k, N \in \N$ and any possible points $a_{k}=\mathbf{v}_{1:N+k}, \mathbf{z}_{1:N+k}$ such that $\{\Vb_{1:N}=\mathbf{v}_{1:N+k}, \Zb_{1:N}=\mathbf{z}_{1:N} \}$ entails that $\{K_{v,N}>1\}$ holds (and thus $\Ind(E_{N})=1)$ we have
\begin{equation*}
    {G^{(N+k)}}\big(a_{k} \mid \mathbf{V}_{1:N}=\mathbf{v}_{1:N},\mathbf{Z}_{v,N}=\mathbf{z}_{v,N}\big)= \dfrac{\widetilde{G^{(N+k)}}\big( a_{k} \big)}{\widetilde{G^{(N+k)}} \{ \mathbf{V}_{1:N} = \mathbf{v}_{1:N}, \mathbf{Z}_{v,N} = \mathbf{z}_{v,N}\}},
\end{equation*}
by \eqref{eq: trunc urn} and definition of conditional probability.

To conclude the proof of \eqref{eq: fin Dir tilde pred} we note that, by \eqref{eq: eventually E_n}, there exists a set $\mathcal{T}$ of sequences $(t_{i})_{i=1}^{\infty}$ that are possible realizations of $(T_i)_{i=1}^{\infty}$  such that $\gt\{\mathcal{T}\}=1$ and such that for any sequence $t=(t_{i})_{i=1}^{\infty} \in \mathcal{T}$ there exists a $\bar{N}(t) \in \N$ such that $\{ K_{v,N}(t)>1\}$ holds for any $N \ge \bar{N}(t)$.
Therefore, for any $N \ge \bar{N}(t)$
$$
{G^{(N+k)}}\big(a_{k} \mid \mathbf{V}_{1:N}=\mathbf{v}_{1:N},\mathbf{Z}_{v,N}=\mathbf{z}_{v,N}\big)=\widetilde{G^{(N+k)}}(a_{k} \mid \mathbf{V}_{1:N}=\mathbf{v}_{1:N},\mathbf{Z}_{v,N}=\mathbf{z}_{v,N}),
$$
where $t_i=v_i$ if $v_i=0$ and $t_i=(v_i, z_i)$ if $v_i=1$. 
Thus we proved \eqref{eq: fin Dir tilde pred}. 
\\
\\
\textbf{Case with a Gnedin process in} \eqref{eq: pr V Z}.\\
Similarly to the previous case, note that under $\gt$, $(K_{v, N})_{N}$ is an a.s.\ non-decreasing Markovian sequence of positive integers.
Moreover, by Kingman representation theorem and the fact that $\gt\{M_{v} < \infty\}=1$ we have that $$\gt\{\{K_{v,N} = M_{v}\} \text{ eventually } w.r.t.\ N\}=1.$$
Indeed, the random partition can be thought of as arising from the ties obtained by sampling from a unique discrete probability measure $P_{v}=\sum_{m=1}^{M_{v}} \pi_{m} \delta_{\tilde{\thb}}$ (here distributed as a Gnedin process) and the frequency of the $k$th largest partition block converges almost surely to $k$th largest random weight in $(\pi_{m})_{m=1}^{M_{v}}$ for any $k \in 1,\ldots, M_{v}$.

Note that $\{K_{v,N} = M_{v}\} \subset \{K_{v,N} >1\} \cup \{M_{v}=1\}  \subset \Tr \cup \{M_{v}=1\}$, thus
$$ \gt\{\{K_{v,N} >1\} \cup \{M_{v}=1\} \text{ eventually } w.r.t.\ N\}=1$$
and
\begin{equation} \label{eq: eventually E_n M_n}
  \gt\big\{ \Tr \cup \{M_{v}=1\} \text{ eventually }  w.r.t.\ N \big\} =1.
\end{equation}
To conclude the proof we need to show that, for any $k \in \N$ and any possible set of points $a_{k}=(\mathbf{v}_{1:N+k},\mathbf{z}_{1:N+k})$ 
\begin{equation} \label{eq: MFM tilde pred}
  \widetilde{G}_{VZ}\left\{ \big\{G^{(N+k)}( a_{k}\mid \mathbf{V}_{1:N},\mathbf{Z}_{v,N}) = \widetilde{G^{(N+k)}}(a_{k} \mid \mathbf{V}_{1:N}, \mathbf{Z}_{v,N}) \big\} \cup \{M_{v}=1\} \text{ eventually}	\right\}=1.
\end{equation}

To prove \eqref{eq: MFM tilde pred} we note that, by \eqref{eq: eventually E_n M_n}, there exists a set $\mathcal{T}$ of sequences $(t_{i})_{i=1}^{\infty}$ that are possible realizations of $(T_i)_{i=1}^{\infty}$  such that $\gt\{\mathcal{T}\}=1$ and such that for any sequence $t=(t_{i})_{i=1}^{\infty} \in \mathcal{T}$ there exists a $\bar{N}(t) \in \N$ such that $\{K_{v,N}(t) >1\} \cup \{M_{v}(t)=1\}$ (and thus $\Tr$) holds for any $N \ge \bar{N}(t)$.
Therefore, by \eqref{eq: trunc urn} and definition of conditional probability, for any $N \ge \bar{N}(t)$
$$
{G^{(N+k)}}\big(a_{k} \mid \mathbf{V}_{1:N}=\mathbf{v}_{1:N},\mathbf{Z}_{v,N}=\mathbf{z}_{v,N}\big)=\widetilde{G^{(N+k)}}(a_{k} \mid \mathbf{V}_{1:N}=\mathbf{v}_{1:N},\mathbf{Z}_{v,N}=\mathbf{z}_{v,N}),
$$
where $t_i=v_i$ if $v_i=0$ and $t_i=(v_i, z_i)$ if $v_i=1$. 
\end{proof}

\subsection{Proof of Proposition \ref{prop: fEPPF}} \label{sec: proof fEPPF}
The fEPPF in Proposition \ref{prop: fEPPF} is computed via marginalization of the pmf of the GARP in \eqref{eq: pr V Z} over all the quantities that are compatible with the cardinalities $\{c_{1},\ldots,c_{K_{v}}\}$ of $\Psi_{N}$. 

We state a more complete version of Proposition \ref{prop: fEPPF}, now including a statement of the range of the three sums that appear in
\begin{multline*}
  \mathrm{fEPPF}_{K_{N}}^{(N)}(|C_{1}|,\ldots,|C_{K_{N}}|) \propto
  \\
  \sum_{N_{v}} \left\{\binom{N}{N_{v}} p_{v}^{N_{v}} (1-p_{v})^{N-N_{v}} \sum_{K_{v}} \left[ \binom{{ M_{e}}}{K_{N}-K_{v}} \right.\right.\\
	\left.\left.
	\sum_{(n_{k_{1}},\ldots,n_{K_{v}})} \EPPF^{(N_{v})}_{K_{v}} (n_{k_{1}},\ldots,n_{K_{v}})\,	\mathrm{DM}^{(N-N_{v})}_{M_{e}} ((n_{\kv,k_{v^{\prime}}})_{\kv<	k_{v}^{\prime}})	\right]\right\}
\end{multline*}
The first sum runs over $N_{v} \in [N]$ with the restriction that $N_{v}=N$ if $K_{N}\le 2$.
The second sum runs over $K_{v} \in [K_{N}]$ with the restrictions that 
\begin{enumerate}
    \item $K_{v} \ge 2$ if $K_{N} \ge 2$;
    \item $K_{v} <K$ if $N_{v}\ne N$;
    \item $K_{v}=K$ if $N_{v}= N$;
    \item $K_{N} \le N_{v} +$ \mbox{$\min\{M_{e},N-N_{v}\}$}, keeping in mind that $M_{e} \coloneqq \frac{K_{v}(K_{v}-1)}{2}$.
\end{enumerate}
Finally, the last sum runs over $ (n_{1},\ldots,n_{K_{v}})$ where $\sum_{\kv=1}^{K_{v}} n_{\kv} =N_{v}$ and $n_{1},\ldots, n_{K_{v}}$ are distinct elements of $\{c_{1},\ldots c_{K_{N}}\}$ ordered, e.g., by cardinalities.
And the non-zero edge-cluster sizes $n_{\kv,\kvp}$ are the remaining (ordered) elements of $(c_{1},\ldots c_{K_{N}})$ that are not matched with vertex-cluster sizes $n_k$.
\subsection{Proof of Proposition \ref{prop: fin exch no inf}}
\begin{proof}
{\bf Finite exchangeability.}\\[.2cm]
First note that $\Zv=(Z_{v, i})_{i=1}^{N_{v}} \coloneqq (Z_{i}: i \in [N], \, V_{i}=1)$ identifies arbitrarily labeled vertex-clusters (e.g., in order of appearance).
Hence, formally the vector $\Zbv$ and its relabeling are regarded as distinct objects, even though they identify the same vertex-partition.

Moreover, if the edge-clusters are relabelled according to the relabeling of the vertex-clusters this identifies the exact same graph-aligned random partition.

For instance, $(Z_{1}=1, Z_{2}=2,Z_{3}=5,Z_{4}=(1,2))$ entails the same graph-aligned partition as $(Z_{1}=3, Z_{2}=2, Z_{3}=1, Z_{4}=(2,3))$, but a different one than $(Z_{1}=3, Z_{2}=2, Z_{3}=3, Z_{4}=(1,2))$.
A relabeling of $Z_{i}$ which preserves the same graph-aligned random partition does not modify the likelihood distribution $\gn(\yb\mid\Vb, \Zb)$ in \eqref{eq: likelihood}, which is invariant under such a relabeling.

By construction, the graph-aligned random partition \eqref{eq: pr V Z} induced by $(V_{i}, Z_{i})$ is exchangeable, i.e., the joint law is invariant to permutation of the labels $i$.
Note that we cannot state the same argument directly in terms of the pmf of $(V_{i}, Z_{i})$ since we have an arbitrary order of $\Zbv$, i.e., the order of arrival (irrelevant for the graph-aligned random partition) that gives probability zero to permutations of $i$'s that entails a non-increasing sequence of $\Zbv$. 

Since the likelihood of the sample \eqref{eq: likelihood} can be defined as a function of the graph-aligned random partition, we immediately obtain the exchangeability of the sample $(\yb)_{i=1}^{N}$. 

Finally, since the random partition $\Psi_{N}$ can be seen as the marginalization of the graph-aligned random partition, we also have finite exchangeability of $\Psi_{N}$ as also shown via the fEPPF \eqref{prop: fEPPF}.
\medskip

{\bf Lack of projectivity.}\\[.25cm]
To prove that \emph{infinity exchangeability} does not hold we show a simple counterexample where projectivity does not hold. 

We first show the lack of projectivity for the graph-aligned random partition $\gn(\Vb,\Zb)$.
It suffices to note that for a sample of size $N=1$ the probability of assigning an observation to a vertex is 1, i.e., ${G^{(1)}}\{V_{1}=1\}=1$, while it is strictly smaller than 1 for $N=3$, since, by \eqref{eq: pr V Z},
$${G^{(3)}}\{V_{1}=0\}={G^{(3)}}\{V_{1}=0,Z_{1}=(1,2),V_{2}=1,Z_{2}=1,V_{3}=1,Z_{3}=2\}>0.
$$
Next, we show the lack of projectivity for $\yb$. 
The last argument also implies that in a sample of size $N=1$ the marginal density of the observations $\yb_{1}$ can be rewritten as 
\begin{equation}
  \int \Norm (\yb \mid\mub^\ast, \Sigb^\ast) \mathrm{d}\mbox{NIW}(\mub^\ast, \Sigb^\ast \mid {\mub}_{0}, \lambda_{0}, \kappa_{0}, {\Sigb}_{0})
  \label{eq: NIW supp}
\end{equation}
while under $N=3$ it is a mixture of \eqref{eq: NIW supp} and an additional term corresponding to an allocation as an edge:
\begin{equation*}
  \int \Norm (\yb \mid\mub_{1,2}^\ast, \Sigb_{1,2}^\ast) \mathrm{d}{G^{(3)}}(\mub_{1,2}^\ast, \Sigb_{1,2}^\ast),
\end{equation*}
with ${G^{(3)}}(\mub_{1,2}^\ast, \Sigb_{1,2}^\ast)$ characterized by $\mub_{1,2}^\ast = (\mub_{1}^\ast+\mub_{2}^\ast)/2$ and $\Sigb_{1,2}^\ast = f(\mub_{1}^\ast,\mub_{2})$, where $\mub_{1}^\ast$ and $\mub_{2}^\ast$ are independent draws of a generalized Student-T distribution.
This shows that $\yb_{i}$ is not infinitely exchangeable.

Finally, we consider the random partition $\Psi_{N}$.
Note that the probability of observations $i=1,2$ being clustered together in a sample of size 2 (i.e., of a partition with a single cluster), is equal to
\begin{eqnarray*}
  {G^{(2)}}\{\Psi_{2}=\{1, 2\}\} & = & \fEPPF^{(2)}_{1}(2) = \EPPF^{(2)}_{1}(2) > \\
                                & > & {G^{(3)}}\{\Psi_{3}: Z_{1}=Z_{2}\} = \EPPF^{(2)}_{1}(2) \, {G^{(3)}}\{V_{1}=V_{2}=1\}.
\end{eqnarray*}
Thus, in the last expression, the first factor is the probability of having the observations with labels $i=1,2$ in the same cluster given that they are in vertex-clusters, and the second factor is the probability of those two observations being assigned to vertex clusters.
Note that, in the case of $N=2$ the probability of the two observations to be assigned in vertex-cluster is 1.
\end{proof}

\subsection{Proof of Theorem \ref{th: limit de Fin Dir} and Corollary \ref{cor: limit MFM}}
Theorem \ref{th: limit de Fin Dir} in the main manuscript shows that in some cases the predictive distributions of the GARP model eventually (i.e., for a large enough sample size $N$) can be characterized as a projection of the predictive distributions of a limiting infinitely exchangeable model, thus where projectivity holds.

\begin{proof}
\textbf{Proof of Theorem \ref{th: limit de Fin Dir} ($M_{v}$-dimensional symmetric Dirichlet)}\\
\textbf{(Case 1: $M_{v}=1$)}\\
For any $N \in \mathbb{N}$ our proposal degenerates to a single Gaussian model because $\gn$-a.s.\ all the observations are clustered together in a single vertex. 
In such a case it is immediate to check that we have projectivity and \eqref{eq: limit urn dir}, \eqref{eq: limit de Fin Dir} and \eqref{eq: G_inf} hold.
However, this is clearly an uninteresting case from a modeling perspective.
\\
\textbf{(Case 2: $M_{v}>1$)}\\
First, recall that $$\gt\bigg\{\lim_{N \rightarrow \infty} \frac{N_{v, N}}{N} \rightarrow p_{v} \bigg\}=1$$ by the strong law of large numbers.

Recall also that under $\gt$, $(K_{v, N})_{N}$ is an a.s.\ non-decreasing Markovian sequence of positive integers such that for any $N\in\mathbb{N}$, $K_{v, N} \le M_{v}$ and $\gt \big\{K_{v, N} = \min(N,M_{v})\big\} >0$ and it can be computed from \eqref{eq: prior V_i}-\eqref{eq: Gibbs urn}.

Moreover, by Kingman's representation theorem (see Kingman, 1978 and Theorem 14.7 in Ghosal and van der Vaart, 2017) the random partition can be thought of as arising from the ties obtained by sampling from a unique discrete probability measure $P_{v}=\sum_{m=1}^{M_{v}} \pi_{m} \delta_{\tilde{\thb}}$ (we know that is $M_{v}$-symmetric Dirichlet distributed) and the frequency of the $k$th largest partition block converges almost surely to $k$th largest random weight in $(\pi_{m})_{m=1}^{M_{v}}$ for any $k \in \{1,\ldots, M_{v}\}$.
Therefore, together with the assumption that $M_{v}$ is finite $\gt\{\lim_{N \rightarrow \infty} K_{v, N} = M_{v}\}=1$. 
Thus, since $K_{v, N}$ are random integers,
\begin{equation}\label{eq: K_v M_v supp}
\gt\left\{\{K_{v, N}=M_{v} \} \text{ eventually w.r.t.\ } N\right\}=1.
\end{equation}

Note also that
$$
\{K_{v, N}=M_{v}\} \subset \Tr. 
$$
Thus, for any $N, k \in \mathbb{N}$ and $ a_{k} = (v_{i},z_{i})_{i=1}^{N+k}$ such that 
$\{\mathbf{V}_{1:N}=\mathbf{v}_{1:N},\mathbf{Z}_{v,N}=\mathbf{z}_{v,N}\}
$
entails that $\{K_{v, N}=M_{v}\}$ holds (and so $\Ind(\Tr)=1$) we have
\begin{align*}
&{G^{(N+k)}}\big((v_{i},z_{i})_{i=1}^{N+k} \mid \mathbf{V}_{1:N}=\mathbf{v}_{1:N},\mathbf{Z}_{v,N}=\mathbf{z}_{v,N}\big)=\\
&\frac{\widetilde{G^{(N+k)}}\big((v_{i},z_{i})_{i=1}^{N+k} \big)}{\widetilde{G^{(N+k)}}\{ \mathbf{V}_{1:N}=\mathbf{v}_{1:N},\mathbf{Z}_{v,N}=\mathbf{z}_{v,N}\}}=\\
&\widetilde{G^{(N+k)}}\big((v_{i},z_{i})_{i=1}^{N+k} \mid \mathbf{V}_{1:N}=\mathbf{v}_{1:N},\mathbf{Z}_{v,N}=\mathbf{z}_{v,N}\big),
\end{align*}
by definition of conditional probability and \eqref{eq: trunc urn}.

Note that, for any $N \in \N$, $K_{v, N}=M_{v}$ entails that $K_{v, N+k}=M_{v}$ and $M_{e,N+k}= \Mep \coloneqq \frac{M_{v}(M_{v}-1)}{2}$ for any $k=0,1,\ldots$.
Therefore, by definition of $\widetilde{G^{(N)}}$ and the fact that $\{\mathbf{V}_{1:N}=\mathbf{v}_{1:N},\mathbf{Z}_{v,N}=\mathbf{z}_{v,N}\}$ entails that $\{K_{v,N}=M_{v}\}$ and $\{M_{e, N} =\Mep \}$ hold, for any $k \in \N$ we have
\begin{align*}
  &\widetilde{G^{(N+k)}}\big((v_{i},z_{i})_{i=1}^{N+k} \mid \mathbf{V}_{1:N}=\mathbf{v}_{1:N},\mathbf{Z}_{v,N}=\mathbf{z}_{v,N}\big)=\\
  &\frac{\ginf_{N+k}\big((v_{i},z_{i})_{i=1}^{N+k} \big)}{\ginf_{N+k}\{ \mathbf{V}_{1:N}=\mathbf{v}_{1:N},\mathbf{Z}_{v,N}=\mathbf{z}_{v,N}\}}=\\
  & \ginf_{N+k}\big((v_{i},z_{i})_{i=1}^{N+k} \mid \mathbf{V}_{1:N}=\mathbf{v}_{1:N},\mathbf{Z}_{v,N}=\mathbf{z}_{v,N}\big),
\end{align*}
where $\ginf_{N+k}$ refers to the pmf of $\mathbf{V}_{1:N+k},\mathbf{Z}_{1:N+k}$ defined in \eqref{eq: G_inf}. 
We now explicitly call such law $\ginf_{N+k}$ (i.e., with the subscript) to stress the dimension to show that $(G_{N}^{(\infty)})_{N \in \mathbb{N}}$ are indeed Kolmogorov consistent and can be seen as the projection of the law of a stochastic process $\ginf$.

To conclude the proof of \eqref{eq: limit urn de Fin} recall that by \eqref{eq: K_v M_v supp}, there exists a set $\mathcal{T}$ of sequences $(t_{i})_{i=1}^{\infty}$ that are possible realizations of $(T_i)_{i=1}^{\infty}$  such that $\gt\{\mathcal{T}\}=1$ and such that for any sequence $t=(t_{i})_{i=1}^{\infty} \in \mathcal{T}$ there exists a $\bar{N}(t) \in \N$ such that $\{K_{v, N}(t)=M_{v} \}$ (and thus also $\{M_{e, N}(t) =\Mep \}$) holds
for any $N \ge \bar{N}(t)$.
Therefore, for any $N \ge \bar{N}(t)$
$$
{G^{(N+k)}}\big(a_{k} \mid \mathbf{V}_{1:N}=\mathbf{v}_{1:N},\mathbf{Z}_{v,N}=\mathbf{z}_{v,N}\big)=\ginf_{N+k}(a_{k} \mid \mathbf{V}_{1:N}=\mathbf{v}_{1:N},\mathbf{Z}_{v,N}=\mathbf{z}_{v,N}),
$$
where $t_i=v_i$ if $v_i=0$ and $t_i=(v_i, z_i)$ if $v_i=1$. 

To check the projectivity of $(\ginf_{N})_{N}$ we note that for any $N \in \mathbb{N}$ and possible values $(v_{i}, z_{i})_{i \in [N]}$
\begin{align*}
    \begin{split}
        \ginf_{N}((v_{i}, z_{i})_{i \in [N]}) &= p_{v}^{N_{v}} \EPPF_{M_{v}}^{(N_{v})}(n_{1},\ldots,n_{K_{v}}\mid \alpha, \sigma)/K_{v}! \\
        &(1-p_{v})^{N_{e}} \DM_{M_{v}(M_{v}-1)/2}^{(N_{e})}((n_{\kv,\kvp})_{\kv< \kvp} \mid \beta/M_{e})\\
        &= \sum_{v_{N+1},z_{N+1}} \ginf_{N+1}((v_{i}, z_{i})_{i \in [N]})=\ginf((V,Z)_{i \in [N]}).
    \end{split}
\end{align*}
The second and third equalities hold by projectivity of the EPPF and DM (where the sum is over all possible values of $v_{N+1}, z_{N+1}$).
We denote by $\ginf$ the infinite-dimensional GARP defined via such Kolmogorov consistent finite-dimensional distributions.

From $\ginf$ (and its Kolmogorov consistent finite-dimensional) we derive the urn schemes in \eqref{eq: limit urn dir} via the definition of conditional probability. The ratio boils down to \eqref{eq: limit urn dir} thanks to the product form of the EPPF and of the DM.

Finally, note that via the characterization of the EPPF and DM in terms of discrete random probabilities (see e.g., Section \ref{sec: discr supp}), the induced law on $(\thb_{i})_{i=1}^{N}$ can thus be characterized by first sampling $V_{i} \iidsim \text{Bern}(p_{v})$ and $\thb_{i} \mid P_{v}, V_{i}=1\indsim P_{v} \coloneqq \sum_{m=1}^{M} \pi_{m} \delta_{\tilde{\thb}_{m}}$ and $\thb_{i} \mid P_{e}, V_{i}=0 \indsim P_{e} \coloneqq \sum_{k<\kvp<M} \pi_{\kv,\kvp} \delta_{\tilde{\thb}_{\kv,\kvp}}$.
Thus we derive \eqref{eq: limit de Fin Dir} marginalizing with respect to $\mathbf{V}$ and by the uniqueness of the directing measure.
\\
\\
\text{\textbf{Proof of corollary \ref{cor: limit MFM}}}\\
First, we write explicitly the statement of Corollary \ref{cor: limit MFM}.
\begin{Cor}[Corollary 1 of the main manuscript]
  Under the {\rm{GARP}} with a Gnedin process (Example \ref{ex: GP}) in \eqref{eq: pr V Z} there exists a finite random sample size $\bar{N}$ such that for any $N>\bar{N}$ the predictive distributions under the proposed GARP model given $M_{v}$ are $\gt$-a.s.\ equal to the predictive distributions given $M_{v}$ under a Kolmogorov consistent $\ginf$, i.e., for any possible sequence of sets of points $(a_{k})_{k\in\N}$, with $a_{k}= \mathbf{v}_{1:N+k}, \mathbf{z}_{1:N+k})$ 
  \begin{equation*}
    \gt\left\{ \big\{\ginf_{N+k}( a_{k} \mid \mathbf{V}_{1:N},\mathbf{Z}_{1:N}, M_{v}) = G^{(N+k)}( a_{k} \mid \mathbf{V}_{1:N},\mathbf{Z}_{1:N},M_{v})\; \forall\,k \big\} \text{ eventually}\right\}=1.
  \end{equation*}
  Moreover, $\ginf(\cdot \mid \mathbf{V}_{1:N},\mathbf{Z}_{1:N}, M_{v})$ can be characterized by the urn scheme in \eqref{eq: limit urn dir} and $\ginf(\cdot \mid M_{v})$ by the pmf \eqref{eq: G_inf} and by an exchangeable sequence with directing measure being the law of $P\mid M_{v}$ as in \eqref{eq: limit de Fin Dir}. 
  Finally, $\ginf(M_{v}=m)= \gt(M_{v}=m)= \frac{\gamma (1-\gamma)_{m-1}}{m!}$.
\end{Cor}

Note that $\gt$-a.s.\ $M_{v} \in \mathbb{N}$ and that for any realization of $M_{v}=m \in \mathbb{N}$ we are back to the finite symmetric Dirichlet GARP and thus the result follows from Theorem \ref{th: limit de Fin Dir}.
\end{proof}

\section{Software, Runtime, etc.} \label{sec: software supp} 

The results reported in this article are based on 10,000 MCMC iterations with the initial 5,000 iterations discarded as burn-in. 
The remaining samples were further thinned by an interval 2. 
We programmed everything in \texttt{R}.
The analyses are performed with a Lenovo ThinkStation P330 with 16Gb RAM (Windows 10), using a \texttt{R} version 4.2.3.
The MCMC algorithm takes 29.8 minutes.

\section*{References}
Binder, D.\ A.\ (1978). Bayesian cluster analysis. \emph{Biometrika}, \textbf{65}, 31--38.\\
\vspace*{-0.5cm}
\\
Dahl, D.\ B., Johnson, D.\ J., and M\"uller, P.\ (2022a). Salso: search algorithms and loss \\
\hspace*{0.3cm}
functions for Bayesian clustering. \emph{R package version 0.3.29.}
\\
\vspace*{-0.5cm}
\\
Dahl, D.\ B., Johnson, D.\ J., and M\"uller, P.\ (2022b). Search algorithms and loss functions\\
\hspace*{0.3cm} for Bayesian clustering. \emph{J.\ Comput.\ Graph.\ Stat.,} \textbf{31}, 1189--1201.\\
\vspace*{-0.5cm}
\\
Franzolini, B.\ and Rebaudo, G.\ (2024). Entropy regularization in probabilistic clustering.\\  \hspace*{0.3cm} \emph{Stat. Methods Appt.,} \textbf{in press.}\\
\vspace*{-0.5cm}
\\
Ghosal, S.\ and van der Vaart, A.\ (2017). \emph{Fundamentals of Nonparametric Bayesian \\
\hspace*{0.3cm} Inference.} Cambridge Univ.\ Press.
\\
\vspace*{-0.5cm}
\\
Kingman, J.\ F.\ C.\ (1978). The representation of partition structures. \emph{J.\ London Math.\ \\
\hspace*{0.3cm} Soc.,} \textbf{18}, 374-380.
\\
\vspace*{-0.5cm}
\\
Meil\u{a}, M.\ (2007). Comparing clusterings--an information based distance. \emph{J.\ Multivar.\ \\ 
\hspace*{0.3cm} Anal.,} \textbf{98}, 873–895.\\
\vspace*{-0.5cm}
\\
Neal, R. M. (2000). Markov chain sampling methods for Dirichlet process mixture models.\\
\hspace*{0.3cm} \emph{J.\ Comput.\ Graph.\ Stat.,} \textbf{9}, 249–265.
\\
\vspace*{-0.5cm}
\\
R Core Team (2021). R: A language and environment for statistical \mbox{computing.} \\ 
\hspace*{0.3cm} \emph{R Foundation for Statistical Computing}, Vienna, Austria.
\\
\vspace*{-0.5cm}
\\
Robert, C.\ P., and Roberts, G.\ (2021). Rao--Blackwellisation in the Markov Chain Monte\\ 
\hspace*{0.3cm} Carlo Era. \emph{Int. Stat. Rev.}, \textbf{89}, 237--249.
\\
\vspace*{-0.5cm}
\\
Teh, Y.\ W., Jordan, M. I., Beal, M.\ J., and Blei, D.\ M.\ (2006). Hierarchical Dirichlet\\
\hspace*{0.3cm} processes. \emph{J.\ Am.\ Stat.\ Assoc.,} \textbf{101}, 1566–1581.
\\
\vspace*{-0.5cm}
\\
Wade, S.\ and Ghahramani, Z.\ (2018). Bayesian cluster analysis: point \mbox{estimation} \\ 
\hspace*{0.3cm} and credible balls. \emph{Bayesian Anal.,} \textbf{13}, 559–626.

\end{document}